\documentclass[pdflatex,sn-mathphys-num]{sn-jnl}%  Math  and  Physical  Sciences  Numbered  Reference  Style
\usepackage{graphicx}%
\usepackage{multirow}%
\usepackage{amsmath,amssymb,amsfonts}%
\usepackage{amsthm}%
\usepackage{mathrsfs}%
\usepackage[title]{appendix}%
\usepackage{xcolor}%
\usepackage{textcomp}%
\usepackage{manyfoot}%
\usepackage{booktabs}%
\usepackage{algorithm}%
\usepackage{algorithmicx}%
\usepackage{algpseudocode}%
\usepackage{listings}%
\usepackage{comment}
\usepackage{adjustbox}

\theoremstyle{thmstyleone}%
\newtheorem{theorem}{Theorem}%    meant  for  continuous  numbers
\newtheorem{proposition}[theorem]{Proposition}%  

\theoremstyle{thmstyletwo}%
\newtheorem{remark}{Remark}%

\theoremstyle{thmstylethree}%

\begin{document}

\title[Article  Title]{Phase  transition  in  a  long-memory  log-Gaussian  Cox  process}

%%=============================================================%%
%%  GivenName	->  \fnm{Joergen  W.}
%%  Particle	->  \spfx{van  der}  ->  surname  prefix
%%  FamilyName	->  \sur{Ploeg}
%%  Suffix	->  \sfx{IV}
%%  \author*[1,2]{\fnm{Joergen  W.}  \spfx{van  der}  \sur{Ploeg}  
%%    \sfx{IV}}\email{iauthor@gmail.com}
%%=============================================================%%

\author*[1]{\fnm{Masato}  \sur{Hisakado}}\email{hisakadom@yahoo.co.jp}

\author[2]{\fnm{Shintaro}  \sur{Mori}}\email{shintaro.mori@gmail.com}
%\equalcont{These  authors  contributed  equally  to  this  work.}

\affil[1]{\orgaddress{\street{Kadoma-cho},  \city{Kanazawa},  \postcode{9201192},  \country{Japan}}}

\affil[2]{\orgaddress{\street{Bunkyo-cho  3},  \city{Hirosaki},  \postcode{0358651},  \country{Japan}}}

%%==================================%%
%%  Sample  for  unstructured  abstract  %%
%%==================================%%

\abstract{We  study  a  stochastic  point  process  with  power-law  temporal  correlations  driven  by  hidden  variables.
%%%%%%%%%%%%%%%%%%%%%%
We  show  that  a  
generalized  Merton  type  model  under  an    exponential-tail  asset  assumption  ---  
obtained  by  replacing  the  Gaussian  cumulative  distribution  
function  with  a  logistic  CDF---  
together  with  an  appropriate  double-scaling  limit,
  converges  to  a  log-Gaussian  Cox  process  
(LGCP)  with  log-normal  intensity.
%%%%%%%%%%%%%%%%%%%%%%%%%
%%%%%%%%%%%%%%%%%
The  resulting  LGCP  exhibits  a  phase  transition  at  the  critical  power  index  $\gamma=1$.
This  transition  separates  regimes  of  short-memory  dynamics  from  long-memory  behavior  characterized  by  anomalous  diffusion.  We  further  demonstrate  that  temporal  correlations  persist  even  in  the  Poisson  limit  when  the  scaling  is  properly  defined,  in  contrast  to  conventional  Poisson  convergence  where  memory  effects  vanish.  We  also  compare  this  LGCP    with  self-exciting  processes  such  as  Hawkes  processes,  highlighting  fundamental  differences  in  correlation  structure  and  extreme-event  behavior.  
The  theoretical  results  are  illustrated  using  credit  
risk  time  series,  and  empirical  estimation  of  the  temporal  
correlation  parameter  from  historical  default  data    provides  evidence  
for  long-memory  behavior  in  pre-1980  credit  portfolios.
%%%%%%%%%%%%%%%%%%%%%%%%%%%%%%%%%%%%%%%
}


%%\pacs[JEL  Classification]{D8,  H51}

%%\pacs[MSC  Classification]{35A01,  65L10,  65L12,  65L20,  65L70}

\maketitle

%\newpage
\section{1.    Introduction}

Anomalous    diffusion    is    one    of    the    most    interesting    topics
in    sociophysics    and    econophysics    \cite{galam,galam2,Man}.
Models    describing    such    phenomena    have    a    long    memory    \cite{Bro,W2,G,M,hod,hui,sch}    and
exhibit    several    types    of    phase    transitions.
In    our    previous    work,    we    investigated    voting    models    for    an    information    cascade    
\cite{Mori,Hisakado2,Hisakado4,Hisakado5,Hisakado6}.
This    model    has    two    types    of    phase    transitions.
The    first        is    the    information    cascade    transition,    which    is    similar    to    the
    phase    transition    of    the    Ising    model    
    that    shows    whether    a    distribution    converges    or    not.
The    other    is    the    convergence    transition    of    the
super-normal    diffusion    that    corresponds    to    an    anomalous    diffusion.
In    this    paper,    we    discuss    the    latter    transition    in    the    limit    of        the    Poisson    process.

In    recent    years,    several    studies    have    been    conducted    regarding    the    time    series    of    financial    markets    from    the    perspective    of    econophysics    \cite{hod,Sor,Kut,Kut2,Taka,Kwe}.
The    important    properties    of    these    data    are    particularly    the    fat    tailed    distribution    of    returns,    long    memory    in    volatility,        and    multi-fractal    nature.
Market    data    of    stock    prices    and    foreign    exchange    have    been    used    in    most    of    these    studies.
The    long    memory        of    volatility    is    known    as    volatility    clustering    \cite{Kwe}    and    
affects    the    risk    management        and
especially        the    calculation    of    Value        at    Risk
(VaR).
This    corresponds    to    the    temporal    correlation        of    the    time    series.
The        temporal    correlation    of    exponential    decay        is    the    short    and    the    power    decay    is    the    medium    or    long    memories.
In    fact,        most    of    the    assets    have    a    short    memory,    but
it    is    known    that    some        assets    have        long    and    medium    memories.

We    study    a    Bayesian    estimation    method    using    the
Merton    model.    Under    normal    circumstances,    the    Merton
model    incorporates    default    correlation    by    
the    correlation    of    asset    price    movements    (asset    correlation),    which
is    used    to    estimate    the    probability    of    default    (PD)    and    the    correlation.
A    Monte    Carlo    simulation    is    an    appropriate    tool    to    estimate    the
parameters,    except    under    the    
limit    of    large    homogeneous    portfolios    
\cite{Sch}.    
In    this    case,    the    distribution    becomes    a    Vasicek    distribution
that    can    be    calculated    analytically    \cite{V}.    

In    our    previous    study,
we    discussed    parameter    estimation    using    a    beta-binomial    distribution    with    default    correlation    and    considered    a    multi-year    case    with    a    temporal    correlation    \cite{Hisakado6}.
    In    Poisson        limit,    we    can    obtain    the    Self-exciting    negative    binomial    distribution    (SE-NBD)    process    and    Hawkes    process.
In    this    study,        we    consider    the    
Poisson    limit    of    the    Merton    model    process    with        temporal    correlation.
We  consider  a  generalized  Merton  framework  with  an  exponential-tail  asset  cumulative  distribution  function  (CDF).
Using        a    logistic    function        instead    of    a    normal    distribution,    we    obtain        the    Poisson    process    with    the    log    normal    intensity    function,        Log    Gaussian    Cox    processes    (LGCP)\cite{cox}.
To    derive    an    analytically    tractable    Poisson    limit,    we    replace    the    normal        CDF        in    the    Merton    model    with    a    logistic    CDF.    Because    the    logistic    decays    exponentially    in    the    tail    while    the    Gaussian    CDF    decays    like    $\exp(-x^2/2)$,    this    substitution    enables    a    log-normal    intensity    in    the    double-scaling    limit    (see    Appendix    \ref{Ac}.    for    details).
%%%%%%%%%%%%%%%%%%
More  generally,  any  asset  distribution  with  an  exponential  
tail  ---  such  as  the  generalized  hyperbolic  distribution  
\cite{Fre}  ---  yields  the  same  LGCP  limit  under  the  
Merton-specific  double  scaling  
$\lambda_0  =  Np'^{1/\sqrt{1-\rho_A}}$,  
whose  exponent  is  dictated  by  the  asset  correlation  $\rho_A$  
of  the  Merton  factor  structure.  Under  the  standard  Poisson  
limit  $\lambda_0  =  Np$,  the  dependence  on  the  latent  variable  
$y_t$  vanishes  and  all  temporal  correlations  are  lost.  
The  Merton-specific  scaling  is  therefore  the  mechanism  by  
which  temporal  correlations  survive  the  Poisson  limit.
%%%%%%%%%%%%%%%%%%%%%

In    this    model    the    super-normal    transition    is    confirmed    when    the
temporal    correlation    follows    a    power    law.
When    the    power    index    is    less    than    or    equal    to    one,    the    diffusion    is    anomalous    faster    diffusion.
It    is    the    super    phase.    This    phase    transition    is    called    the
super-normal    transition.
We    discuss    a    phase    transition    when    we    use    the    Merton    model    and    compare        it    to    SE-NBD        and    Hawkes    models.        
%%%%%%%%%%%%%%%%%%%%%%%%%%%%%%%%%

The  present  paper  makes  three  new  contributions.  First,  we  
establish  a  new  pathway  from  a  Merton-type  credit  risk  model  
to  the  LGCP  via  the  double  scaling  limit,  which  appears  to  
be  new  in  the  literature.  Second,  we  show  that  the  resulting  
LGCP  exhibits  a  super-normal  phase  transition  at  $\gamma  =  1$,  
with  direct  implications  for  the  convergence  of  
probability-of-default  estimators  in  credit  risk  management.  
Third,  we  provide  empirical  evidence  from  historical  default  
data  that  pre-1980  credit  portfolios  operated  in  the  
super-normal  regime  ($\gamma  \leq  1$).

%%%%%%%%%%%%%%%%%%%%%%%%%%%%%%
The    remainder    of    this    paper    is    organized    as    follows.
In    Section    \ref{2},    we    introduce    the    stochastic    process    
  with  hidden  value    and    obtain    the    Poisson    process    in    the        limit.
    We    discuss    the    super-normal    phase    transition    using        the    impact    analysis.
    In    Section    \ref{3}    we    calculate        the    variance    directly    and    confirm    this    phase    transition.
In    Section    \ref{4},    we    describe    the    application    of    the    Bayesian    estimation    approach    to    the    empirical
data    of    default    history    using    the    Merton    model    and    confirm    its    parameters.    
The  main  results  are  summarized  in  proposition  form  in  Section~\ref{5}.  
Finally,    the    conclusions    are    presented    in    Section    6.

\section{  Asset        and    default    correlations}
\label{2}
\subsection{  Introduction    of    the    Model}    
In    this    section,    we    consider        the    time    series    of    a    stochastic    process
using    the    Merton    model    which    has    hidden    variables        \cite{Mer}.    
We    study    annual    default    counts    in    a    portfolio    of    
$N$    obligors.    
The    observable    quantity    is    $N_t$,    the    number    of    defaults    in    $t$-th    term.    
The    latent    variable    $y_t$    represents    the    macroeconomic    environment    and    induces    temporal    dependence.
We    take    the    limit        of    the    process,    Eq.(\ref{condition}),        and    obtain    the    
Poisson    process    with    the    log    normal    intensity    function.

%%%%%%%%%%%%%%%%%    Temporal    correlation
In    the    first    step    we    define        hidden    variables    of    Merton    model.
Normal    random    variables,    $y_t$,    are    hidden    variables    that    explain
the    state        of    the    economy,    and    $y_t$    affects    all    obligors    in    the    $t$-th        term.
Through    $y_t$,        temporal    correlation    affects    the    obligors.
This    is    the    hidden    variable,    which    we    cannot    observe.
To    introduce    the    temporal    correlation    
from    different    terms,    let    $\{y_t,1\le    t    \le    T\}$    be    the    time    series    of    the    stochastic
        variables    of    the    correlated    normal    distribution    with    the    following
    correlation    matrix:
\begin{equation}
\Sigma\equiv\left(
        \begin{array}{ccccc}
        1    &        d_1    &    d_2    &\cdots&    d_{T-1}    \\
        d_1        &        1    &        d_1    &    \ddots&\vdots    \\
        \ddots    &    \ddots    &    \ddots        &    \ddots&\ddots        \\
    \vdots    &    \ddots    &    \ddots        &    \ddots    &d_1\\
        d_{T-1}&        \cdots    &        d_2&        d_1        &    1        \\
        \end{array}
        \right),
        \label{matrix}
\end{equation}
    where    $(y_1,\cdots,y_T)^T\sim    \mbox{N}_{T}(0,\Sigma)$.

In    this    study,    we    consider    two    cases    of    temporal    correlation:    exponential    decay,    $d_{i}=\theta    ^i,0\le
\theta\le    1$,
and    power    decay,    $d_{i}=1/(i+1)^{\gamma},\gamma\ge    0$.
The    exponential    decay    corresponds    to    short    memory,    and    the    power    decay    corresponds    to
intermediate    and    long    memories    \cite{Long}.
In    the    case    of    exponential    decay,
we    can    write
\begin{equation}
        y_{t+1}=\theta    y_t+\sqrt{1-\theta    ^2}\xi_{t+1},
\label{y}
\end{equation}
where    $\xi_t\sim    \mbox{N}(0,1)$,    i.i.d.,        and    the    correlation    between    $y_t$    and    $y_{t+1}$    is    $\theta$    where    $\theta$    is    the    exponential        temporal    correlation.
The    first    term        corresponds    to    the    temporal    correlation    decay.

%%%%%%%%%%%%%%%        Asset    correlation
As        second    step    we    explain    the    Merton    model    at    $t$-th    term    using    the    hidden    variables.
We    assume    that    the    number    of    obligors    in    the    $t$-th    term
is    constant    and    denote    it    as    $N$.    
The    asset    correlation,    $\rho_A$,    is    the    parameter    that    describes    the    
correlation    between    the    value    of    the    assets    of    the    obligors    in    the    same    term.
We    consider    the    $i$-th    asset    value,    $U_{it}$,    at    term        $t$,    to    be
\begin{equation}
        U_{it}=\sqrt{\rho_A}y_t+\sqrt{1-\rho_A}\epsilon_{it},
        \label{rho}
\end{equation}
where    $\epsilon_{it}\sim    \mbox{N}(0,1)$    is    i.i.d.
The    first    term    is        the    economic    effect    and    the    second    term    is    the    individual    obligor    effect.
By    this    formulation,    the        correlation    of    $U_{it}$    is,    
\begin{equation}
        E[U_{it}    U_{jt}]=
\left\{    
\begin{array}{ll}
1    &    (i=j)    \\
\rho_A    &    (i\neq    j)
\end{array}    \right.
\end{equation}
\begin{comment}
$y_t$        is    the    economic    cycle    variable    or    systematic    factor    which    is    hidden    variable.    The    economic    cycle    variable        has    a    temporal    correlation    as    Eq.(\ref{matrix}).
\end{comment}
Here,    we    consider        only    the    positive    correlation,    $\rho_A$,        because    of        Eq.(\ref{rho}).

The    discrete    dynamics    of    the    process    are    described    by
\begin{equation}
        X_{it}=1_{U_{it}    \leq    Y},
\label{prpcess}
\end{equation}
where    $Y$    is    the    threshold        and    $1    \leq    i    \leq    N$.
When    $X_{it}=1    (0)$,    the    $i$-th    obligor    in    the    $t$-th    term
is    default    (non-default).
Eq.(\ref{prpcess})    corresponds    to    the    conditional    default    probability    for    $y_{t}=y$.
It    is    the    probability        the    $i$-th        obligor        value,            $U_{it}$    is    below    the    threshold    $Y$        at    $t$    \cite{Mer}.
The    barrier    option    type    was    introduced    in    \cite{Bla}
In    this    study,    we    extend    the    model    to    the    multi-term    model    with    the    temporal    correlation    \cite{Tas,Tas2}.
In    these    models,    the        clustering    of    defaults    were    discussed    \cite{daf,az}.

Here    we    consider    the    conditional    default    probability    at    $t$-th    term.
Note    that        the    index,    $i$,    is    removed    because    we    do    not    distinguish    each    obligor    in    the    portfolio.
\begin{eqnarray}
G(y)&\equiv&    \mbox{P}(X_{t}=1|y_{t}=y)=\mbox{P}(U_{t}<Y|y_{t}=y),
\nonumber    \\
&=&
\mbox{P}    [\sqrt{\rho_A}    y_t+\sqrt{1-\rho_A}\epsilon_{t}<Y|y_t=y]
\nonumber    \\
&=&
\mbox{P}    [\epsilon_{t}<\frac{Y-\sqrt{\rho_A}    y}{\sqrt{1-\rho_A}}|y_t=y]
\nonumber    \\
&=&
\Phi    \left(\frac{Y-\sqrt{\rho_A}    y}{\sqrt{1-\rho_A}}\right)
    \label{V}
\end{eqnarray}
where    $\Phi(x)$    is    the    cumulative        normal    distribution,    
$G(y_t)$    is    the    distribution    of    the        conditional    default    probability
at    the    $t$-th    term        in    the    portfolio    with    the    economic    cycle    variable,    $y_t=y$
and    the    average    probability    of    default    (PD)    is    $p'=\Phi(Y)$,    which    is    the    long    term    average    of    PD.
It    is    the    Merton    model    at    the        $t$-th    term.
The    temporal    correlation    is    included    in    $y_t$    by    Eq.(\ref{matrix}).

%It    is    the    original    process.
%%%%%%%%%%%%%%%%%%%%%%
This        is    a        discrete    model        \cite{Tas,    Tas2}.
When    the    power    index,    $\gamma$        of    the    temporal    correlation    is
less    than    one,    the    PD        estimation                converges    slowly    to    the    delta    function    \cite{Hisakado8}.    
In    contrast,    when    the    power    index    is    greater    than    one,    the    convergence    is    
same    as    that    in    the    normal    case.    
We    call    this    phase    transition    the    super-normal    transition.
When    the    distribution    converges    slowly,        the    estimation    of    the
long-run    PD    with    limited    data    becomes    time-consuming.
Therefore,    we    believe    it    is    important    for    the        financial    risk    management.

%%%%%%%%%%%%%%%%%%%%%%%%%%%%%    Portfolio

%%%%%%%%%%%%%%%%555%    Limit    of    the    process
Next,    we    consider    the        limit,    $p'\rightarrow    0$    and    $N    \rightarrow\infty    $    with    the    condition    Eq.(\ref{condition})    of    this    process.
The    purpose    of    the    limit    is    to    obtain    the    process        including    the    correlations    and    confirm    the    phase    transition.    
Here,    we    set    $y_t=y$  and  $N_t$  as  the  number  of  defaults  at  $t$-th  term.
The            distribution    of        the    number    of    the    defaults    in    the    portfolio    which    includes    $N$        obligors    at    the        $t$-th    term        is    
\begin{eqnarray}
        P[N_t=k_t]&=&\int_{-\infty}^{\infty}    \frac{N!}{k_t!(N-k_t)!}
        G(y)^{k_t}
        (1-G(y))^{N-k_t}
        \phi(y)    dy,
        \nonumber    \\
        &=&\int_{-\infty}^{\infty}    \frac{N!}{k_t!(N-k_t)!}
        \Phi    \left(\frac{\Phi^{-1}(p')-\sqrt{\rho_A}    y}{\sqrt{1-\rho_A}}\right)^{k_t}
        (1-\Phi    \left(\frac{\Phi^{-1}(p')-\sqrt{\rho_A}    y}{\sqrt{1-\rho_A}}\right))^{N-k_t}
        \nonumber    \\
        &&  \times    \phi(y)    dy,
        \label{bin}
\end{eqnarray}
where    $\phi(y)$    is    the    normal    distribution,    $1/\sqrt{2\pi}e^{-y^2/2}$    and    
\begin{equation}
N_t=\sum_{i=1}^{N}    X_{it.}
\end{equation}
$N_t$        is    the    number    of    defaults        and    $N_t/N$    is    the    probability    of    defaults    in    $t$-th    term.
This    is    a    counting    process.
The    estimation    of    these    is        important    for    the    management    of    credit    or    default    risk    of    portfolio.
If    we    take    the    large    portfolio    limit    $N\rightarrow    \infty$    we        obtain    the            Vasicek    distribution    \cite{V}    which    is    used    in    Basel    II    \cite{Tas}.

Here    we        use        a    logistic    cumulative    function    instead    of    a    normal    cumulative    distribution    function,
\begin{equation}
        \hat{\Phi}(x)=\frac{1}{1+e^{-\beta    x}},
\label{lf}
\end{equation}
and    $\beta\sim    1.3$.
The    substitution    is    not    intended    as    a    literal    replacement    of    the    classical    Merton    model.
It    should    be    viewed    as    a    stylized    extension    incorporating    heavier-tailed    asset    distributions    than    that    of    the    normal    distribution    \cite{Man}.
The    resulting    exponential    tail    is    what    mathematically    generates    the    LGCP    limit.
We    discuss    them    in    Appendix    \ref{Ac}.

In    what    follows    we    model    the    conditional    default    probability    using    a    logistic    cumulative    distribution    function    (CDF)    rather    than    the    normal        CDF.    
The    logistic    tail    is    exponential,    which    leads    to    a    log-normal    distribution    for    the    Poisson    intensity    under    our    double-scaling    limit.    
The    Gaussian    tail    ($\sim        \exp    (-x^2/2)$)    does    not    permit    the    same    derivation    —    see    Appendix    \ref{Ac}    for    a    full    discussion    and    robustness    remarks.

Applying    this    assumption,    we    obtain    
\begin{equation}
        G(y)=\Phi    \left(\frac{\hat{\Phi}^{-1}(p')-\sqrt{\rho_A}    y}{\sqrt{1-\rho_A}}\right)\equiv    p\sim\frac{1}{1+((1-p')/p')^{1/\sqrt{1-\rho_A}}e^{\beta\sqrt{\rho_A}y/\sqrt{1-\rho_A}}},
        \label{appp}
\end{equation}
We    clarify    the    meaning        of    this    assumption    in    Appendix.C.
We    use    the        reverse    function    of    the    logistic    function,    Eq.(\ref{lf}),
\begin{equation}
Y=\hat{\Phi}^{-1}(p')=    \frac{1}{\beta}\log        \frac{p'}{1-p'}.        
\end{equation}

In    the    case    $p=G(y)\ll1$,    we        obtain,
\begin{equation}
    G(y)=        p'^{1/\sqrt{1-\rho_A}}    e^{-\frac{\sqrt{\rho_A}}{\sqrt{1-\rho_A}}\beta    y},
\label{G}
\end{equation}
where  we  use  the  approximation  $1-p'\approx1$.
%%%%%%%%%%%%%
Here,  $p'=\Phi(Y)$  is  the  unconditional  (long-run  average)  probability  of
default:  it  is  obtained  by  averaging  the
conditional  default  probability  $G(y)$  over  the  macroeconomic  factor  $y$,
and  represents  the  portfolio's  annualized  average  PD  over  the  full
economic  cycle.  Empirically,  annual  PDs  for  investment-  and
speculative-grade  portfolios  are  typically  well  below  $10\%$
(cf.Fig.  \ref{fig:moody-defaults},  so  it  is  natural  to  treat  $p'$  as
a  small  parameter,  $p'  \ll  1$.

%%%%%%%%%%%%%%%

Here,    we    take    the    limit
    $p'\rightarrow    0$    and    $N    \rightarrow\infty    $    with    the    condition,    fixed
    \begin{equation}
    \lambda_0=Np'^{1/\sqrt{1-\rho_A}}.
    \label{condition}
    \end{equation}

This  factorization  of  Eq.(\ref{G})  is  the  mechanism  by  which  the  temporal  correlation,  prescribed  through  the  covariance  structure  of  $\{y_t\}$,  survives  the  double-scaling  limit:  because  
$p'$  and  
$y$  enter  Eq.(\ref{G})  multiplicatively,  the
$y$-dependence  —  and  hence  the  temporal  correlation  —  is  unaffected  by  taking  $p'  \to  0$.  
  Therefore,    by    keeping    Eq.(\ref{condition})
fixed,    the    temporal    correlation    remains    in    the    Poisson    limit.    
%    Using    the    separation    of    $p'$    and    $y$,    we    can    take    the    limit    with    the    temporal    correlation.
    The    relation        between    $p$    and    $p'$    is    given    by        Eq.(\ref{appp})    and    $p    \rightarrow    0$    as    $p'\rightarrow    0$.
\begin{comment}
    It    is    the    double    scaling    limit    among    $p$    and    $N$    with    the    scaling    and    including    the    correlation    $\rho_A$.
The    model    in    this    limit        includes    the    correlation.
\end{comment}
%We        can    observe    the    super-normal    phase    transition        observed    in    the    discrete    model.
    If    we    take    simple        limit        $p',p,    \rho_A    \rightarrow    0$    and    $N    \rightarrow\infty    $    with    the    condition,    $\lambda_0=pN$    fixed,        we        obtain    the        Poisson    process    with    the    intensity    $\lambda_0$    and    the    correlation        disappears.    In    this    case    we        do    not    observe    the    phase    transition.

    We    define    the    number    of    the    defaults    at    the        $t$-th    term,    $\lambda(y_t)=\lambda(t)$    as
    \begin{equation}
        \lambda(y)=    Np'^{1/\sqrt{1-\rho_A}}e^{-\frac{\sqrt{\rho_A}}{\sqrt{1-\rho_A}}\beta    y}=\lambda_0        e^{-\frac{\sqrt{\rho_A}}{\sqrt{1-\rho_A}}\beta    y}
        \label{app}
\end{equation}
The    expected    value    of    $\lambda(y)$    is
\begin{equation}
        \bar{\lambda}=\int_{-\infty}^{\infty}
        \lambda(y)\phi(y)    dy=
        \lambda_0    e^{\frac{\rho_A}{2(1-\rho_A)}\beta^2}
        =\lambda_0        e^{\alpha^2/2},
\end{equation}
where    $\alpha=\frac{\sqrt{\rho_A}}{\sqrt{1-\rho_A}}\beta$.

Here,    we    change    the    variable        from    $y$    to    $\hat{y}=-y$.
The    standard    normal    distribution        has        symmetry    in    $y$,    so    we    can    change    the    variable    without        loss    of    generality.
We        obtain
    \begin{equation}
        \lambda(\hat{y})=    \lambda_0        e^{\frac{\sqrt{\rho_A}}{\sqrt{1-\rho_A}}\beta    \hat{y}}
        =\lambda_0        e^{\alpha    \hat{y}}.
\end{equation}
In        the    exponential    decay    case,        we    can    obtain        another    representation
\begin{equation}
    \lambda(\hat{y}_{t+1})=    \lambda_0    e^{\alpha    \hat{y}_{t+1}}=\lambda_0^{1-\theta}        e^{\sqrt{1-\theta^2}\alpha    \hat{\xi}_{t+1}}\lambda(\hat{y}_t)^{\theta},        
\end{equation}
using        Eq.(\ref{y})        and    $\hat{\xi}_{t}=-\xi_t$.

In    this    limit        
we    can    calculate    Eq.(\ref{bin})
        as
    \begin{eqnarray}
    P[N_t=k_t]&\sim&\int_{-\infty}^{\infty}
    \frac{\lambda(\hat{y})^{k_t}e^{-\lambda(\hat{y})}}{k_t!}
    \phi(\hat{y})
        d\hat{y}
        \nonumber    \\
    &=&\int_{0}^{\infty}
    \frac{\lambda(\hat{y})^{k_t}e^{-\lambda(\hat{y})}}{k_t!}f(\lambda)d
    \lambda,        
    \end{eqnarray}
where
\begin{equation}
        f(\lambda)=\phi(\hat{y})\frac{d\hat{y}}{d\lambda}=\frac{1}{\sqrt{2\pi}\alpha    \lambda}e^{-\frac{(\log    \lambda-\log    \lambda_0)^2}{2\alpha^2}}
        \label{f}
\end{equation}
Then,    the    distribution    of    the    intensity    function    is        the    log    normal    distribution.
This    is        the    Poisson
process    with    the    log    normal    intensity    function,    Log    Gaussian    Cox    processes    (LGCP)    \cite{cox}.
The    expected    value    is    $\bar{\lambda}=\lambda_0    e^{\alpha^2/2}$    and    the    variance    is    $    \bar{V}=\bar{\lambda}^2    (e^{\alpha^2}-1)$.
Note    that    the    expected    value    is    a    function    of    $\rho_A$    in    this    process.
Conversely,    in    the    original    process    the    expected    value    of    the    probability    of    defaults    does    not    depend    on    the    function    of    $\rho_A$.        
In    the    limit    $\rho_A    \rightarrow    0$,    $\bar{\lambda}=\lambda_0$    and    $\bar{V}=0$,    this        yields    the    Poisson    process.
In    the    limit    $\rho_A\rightarrow    1$,    $\alpha\rightarrow    \infty$    and    $f(\lambda)\sim    1/\lambda$    where
$\bar{\lambda}\rightarrow    \infty    $    and    $\bar{V}\rightarrow    \infty$.

\begin{figure}[h]
\begin{center}
\begin{tabular}{c}
%    1
\begin{minipage}{0.5\hsize}
\begin{center}
\includegraphics[clip,    width=7cm]{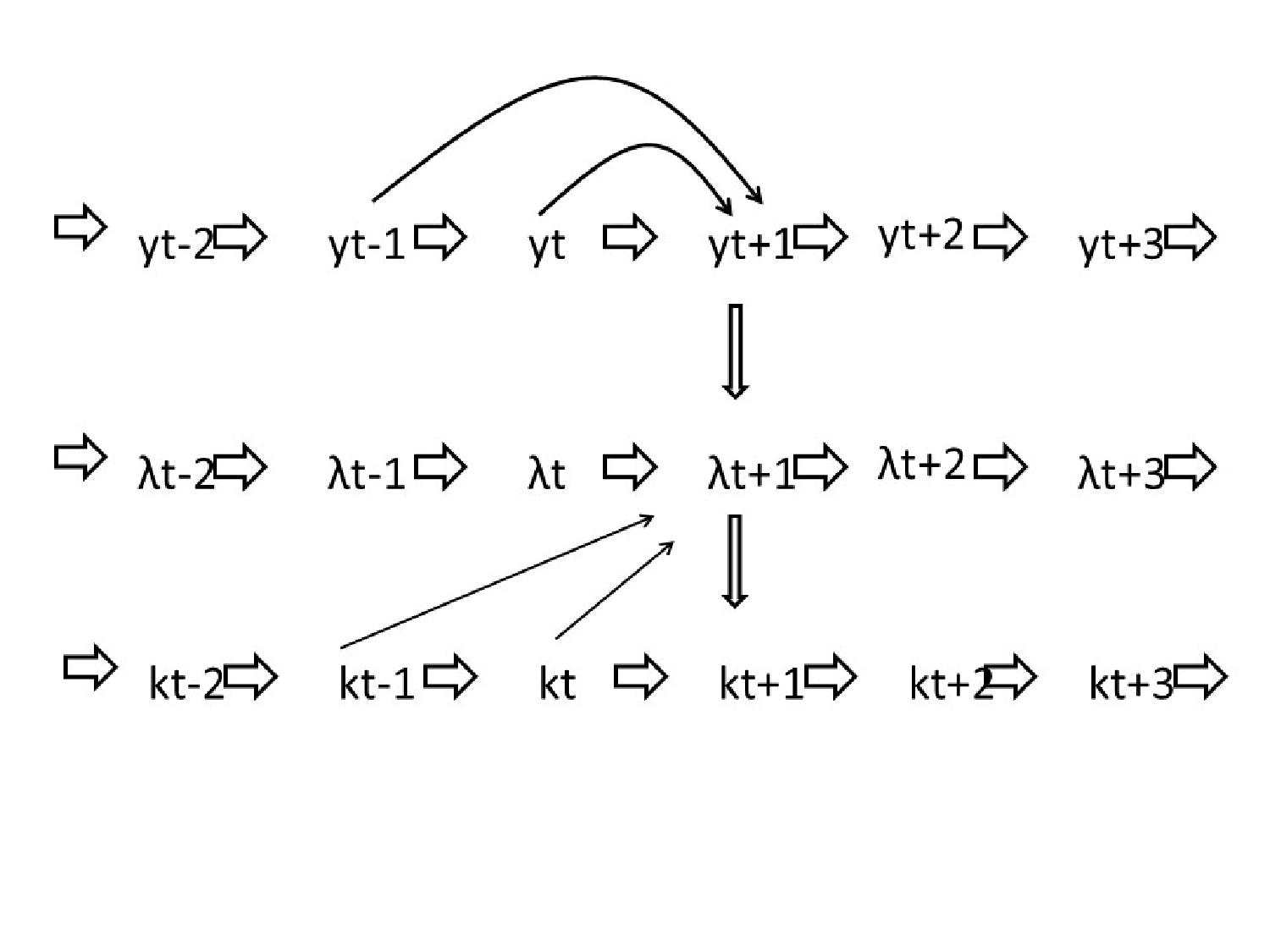}
\hspace{1.6cm}    (a)
\end{center}
\end{minipage}
%    2
\begin{minipage}{0.5\hsize}
\begin{center}
\includegraphics[clip,    width=7cm]{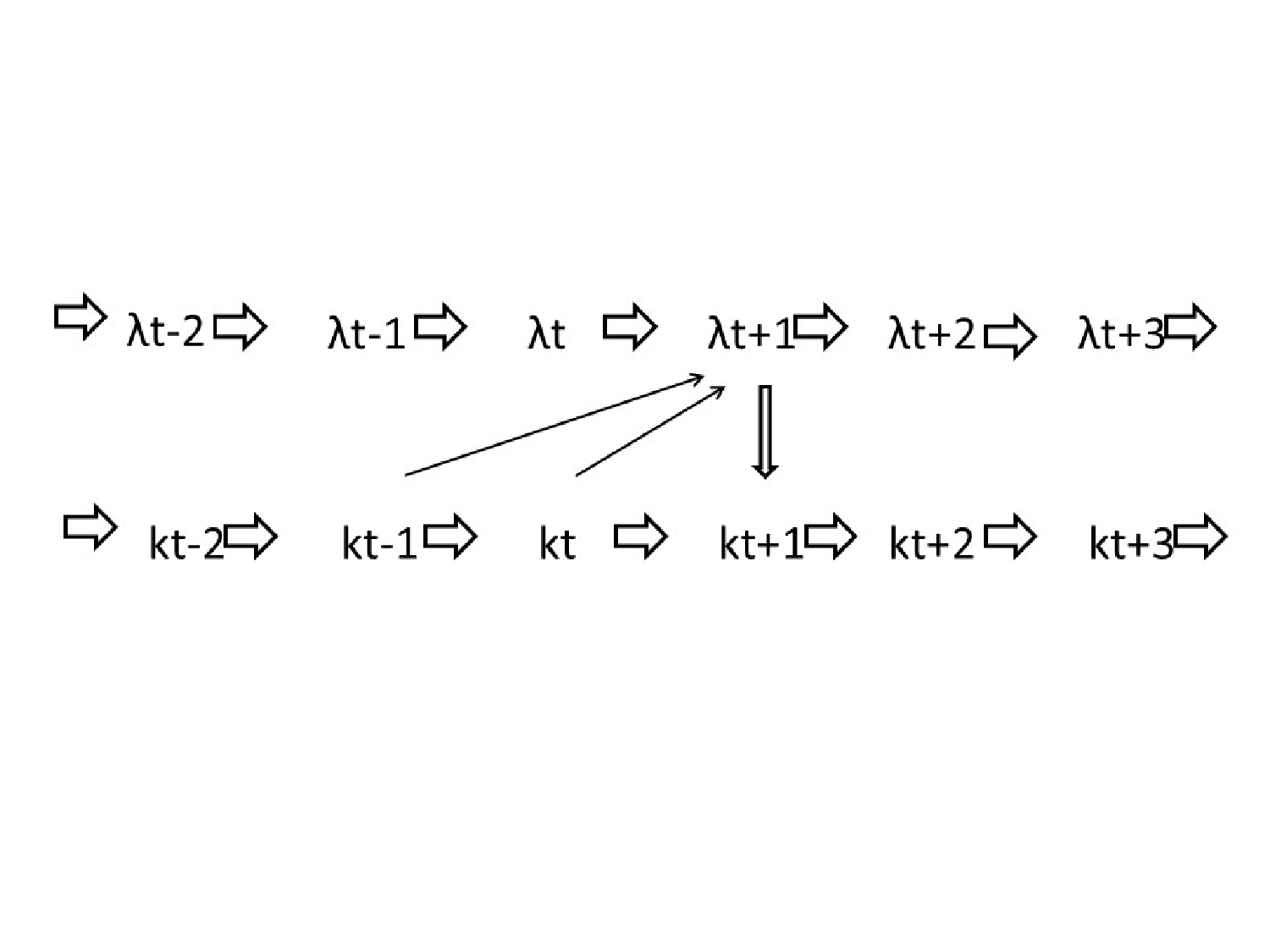}
    \hspace{1.6cm}    (b)
\end{center}
\end{minipage}
%    3
    \end{tabular}
\caption{Relation    among    the    variables        of    (a)    this        process    and    (b)    SE-NBD    and    the        Hawkes    process.    In    (a)    $y_t$    is    the    hidden    process.    The    hidden    process    is    the    status    of    economics    and    used    to    introduce    the    temporal    correlation.    Conversely,        self    excitation    is    used        for    (b).    The    arrows    show    the    effects        of    variables    and    the    double    line    arrows    show    the    calculation    processes.        }
\label{impact}
\end{center}
\end{figure}

    Several    methods        are    available        to    introduce    temporal    correlations.    
    For        the    Hawkes    and        SE-NBD    process    the    self-excitation    is    used    for    the    temporal    correlation.    In    contrast,    in        the    Merton    model    the    temporal    correlations    arise    from    hidden    variables.    
Fig.\ref{impact}    (a)    shows    the    image    of    the    relation    among    the    variables    of    this    process.
For        comparison    we    show    that    of        SE-NBD    and    Hawkes    processes    in    Fig.\ref{impact}    (b).
%We    obtained        three    models        from    urn    model,    the    Poisson    ,    Hawkes,    and        SE-NBD    processes.
%We    confirmed    the    phase    transition    of    urn    model    in    Hawkes    and    SE-NBD    process.    
In    the    simple    limit    we    obtain    the        Hawkes    process,    but    in    the    double    scaling    method    we    obtain        SE-NBD    model.
Both    models        include    the    temporal    correlation.

In    the    model    discussed        in    this    paper    we    obtain    Log    Gaussian    Cox    process    (LGCP)    which    is    the    Poisson    process    with    the    log    normal    intensity    function.
%There    are    several        limits        to    obtain    the    Poisson    process    from    the    discrete    model,    because    the    discrete    model    is        rich.
In    this    process    $y_t$        is        independent    of    the
observable    historical    data    of        $k_t$.
The    temporal    correlation    is    introduced        by        the    time    series    of    $y_t$        where    $y_t$    is    the        hidden    process    which    is    the    correlated    standard    normal    distribution.
It    is    exogenous    factor    driven    temporal    correlation.
$y_t$    affects    the    number    of        defaults    through    the    intensity    function    $\lambda_t$.
In        SE-NBD    and        the    Hawkes    processes,        the    self    excitation    exists        between    the    intensity    function    $\lambda_t$
and    the    number    of    defaults    $k_t$.
Using    self    excitation,    the    temporal    correlation    is    introduced    for    these    models.
The    comparison    of    the    temporal    correlation    from    the    distribution    of    the    intensity    function        is        presented    in    Appendix    \ref{Aa}.    
%In    conclusion,    SE-NBD        and    Hawkes    processes    are    the    positive    feedback    system.    On    the    other    hand,    this    system    is    the    hidden    process    by        temporal    correlation.

The    tail    of    the    LGCP    is    not    the    power    distribution    but    the    log    normal    distribution.    Then,    the    distribution    of    the    maximum    values                belongs    to        the    Gumbel    family.    
The    super-normal    transition    does    not            change    the        distribution.
In    the    strong
correlation    limit    the    tail    of    LGCP    has        the    power    tail.    
On    the    other    hand,    at    the    critical    point,
the    tails    of    Hawkes    and    SE-NBD    process        have        power    distributions    \cite{Hisakado11}.    The    phase    transition    of
Hawkes    and    SE-NBD    is    the    change    from    the    Gumbel    family    to    the    Fréchet    family        by    the    phase    transition    in    the    extreme    value    theory.
This    is    the    difference    between    the    Hawkes,    SE-NBD    models    and        LGCP    with    log    normal    intensity    function    in    the    extreme    value    theory.

%%%%%%%%%%%%%%%%%%%%%    impact    analysis
\subsection{Impact    Analysis}
Next,    we    consider    the    impact    analysis.
This    is    a        study    of    the        effect    of        the    shock    of    the    noise    $\delta$    at    term    $t$.
When        the    shock    is    added    in    $\hat{y}_t$,    we    consider    the        added    expected    value        of    the        intensity    function
from    $t+1$    to    $\infty$.
%%%%%%%%%%%%%%%%%%%%%%

Throughout this paper, we follow the terminology used in the
non-equilibrium statistical physics literature on anomalous
diffusion~\cite{hod,Hisakado2}, where a non-analytic change in the
scaling exponent of a diffusion process at a critical value of a
control parameter is referred to as a ``phase transition,'' by
analogy with equilibrium phase transitions, even though the two
regimes are not thermodynamic phases in the strict sense. We
acknowledge that this usage is not universal: many authors prefer
the more neutral term ``scaling transition'' because what changes
across $\gamma=1$ is a scaling exponent rather than a transition
between distinct equilibrium phases. Concretely, as we establish
quantitatively in Section~\ref{3}, the growth rate of the
cumulative variance of the intensity process changes non-analytically
at $\gamma=1$: it grows linearly in $T$ for $\gamma>1$, but as
$T^{2-\gamma}$ (with a logarithmic correction exactly at $\gamma=1$)
for $\gamma\le1$. It is this non-analytic change in the growth-rate
exponent that we refer to as the super-normal transition. We retain
the term ``phase transition,'' consistent with
Refs.~\cite{hod,Hisakado2,Hisakado11}, while recognizing that the
alternative term ``super-normal scaling transition'' would be equally
appropriate.
%%%%%%%%%%%%%%%%%%%%%%%%%%%%%%%.
\subsubsection{Exponential    decay    case}
First,    we    consider    the    case    of    the        exponential    decay    model    $d_i=\theta^i,\theta\leq    1$.
We        assume        that        the    impact    $\delta$        is    added    at        the    term    $t$    in    the    hidden    variable.
We    add    $\delta$        at    time    $t$    as    an    impact        on        the    intensity    function.
We    evaluate    the    effect    of        the        ratio    of        $\bar{\lambda}_{\delta    \infty}/\bar{\lambda}_{\infty}$.
At    $t\rightarrow    \infty$,        
    $\bar{\lambda}_{\delta    \infty}$    
    is    the        $\bar{\lambda}$    with        the    impact    and        $\bar{\lambda}_{    \infty}$    is    without    the    impact.
    Using    $\bar{\lambda}_{\delta    \infty}/    \bar{\lambda}_{\infty}$,    we    can    estimate    the    effects    of    impact    at        term    $\infty$.
    We    will    confirm    whether    the    effects    of    impact        remain    or    not    at    $t    \rightarrow    \infty$.
    It    is    the    response    function    in    statistical    physics.

We  define  the  cumulative  log-impact  as
\begin{align}
\log  \frac{\bar{\lambda}_{\delta\infty}}{\bar{\lambda}_\infty}  
\equiv  \sum_{s=0}^{\infty}  \mbox{E}\left[\log  \frac{\lambda(\hat{y}_{t+s}  +  \theta^s\delta)}{\lambda(\hat{y}_{t+s})}\right].
\nonumber
\end{align}
Since  $\lambda(\hat{y})  =  \lambda_0  e^{\alpha\hat{y}}$,  each  term  in  the  sum  reduces  to
\begin{align}
\log  \frac{\lambda(\hat{y}_{t+s}  +  \theta^s\delta)}{\lambda(\hat{y}_{t+s})}  
=  \alpha\theta^s\delta,
\nonumber  
\end{align}
which  is  independent  of  $\hat{y}_{t+s}$.  
Therefore,  the  expectation  acts  trivially,  and  summing  over  $s$  yields

\begin{equation}
\log    \frac{\bar{\lambda}_{\delta    \infty}}{\bar{\lambda}_{\infty}}=\mbox{E}[
\log    \frac{\lambda_0    e^{\alpha\delta+\alpha    \hat{y}_t}}
{\lambda_0    e^{\alpha    \hat{y}_t}}
\frac{\lambda_0    e^{\alpha\delta    \theta+\alpha    \hat{y}_{t+1}}}
{\lambda_0    e^{\alpha    \hat{y}_{t+1}}}
\frac{\lambda_0    e^{\alpha\delta    \theta^2+\alpha    \hat{y}_{t+2}}}
{\lambda_0    e^{\alpha    \hat{y}_{t+2}}}
\cdots
]
=\frac{\alpha    \delta}{1-\theta}.
\label{i1}
\end{equation}
In    this    case    the    impact    is    the    finite.

\subsubsection{Power    decay    case}
Next    we    consider        the    following        case,
    $d_i=1/(i+1)^{\gamma},i=1,2,\cdots$,
where    $\gamma\ge    0$    is    the    power    index.
We    consider    the    ratio        for    the    impact    $\delta$    added.
Similarly,  for  the  power  decay  case  $d_i  =  1/(i+1)^\gamma$,  
the  impact  at  lag  $s$  on  $\hat{y}_{t+s}$  is  $d_s  \cdot  \delta  =  \delta/(s+1)^\gamma$,  
giving
\begin{align}
\log  \frac{\lambda(\hat{y}_{t+s}  +  \delta/(s+1)^\gamma)}{\lambda(\hat{y}_{t+s})}  
=  \frac{\alpha\delta}{(s+1)^\gamma},
\nonumber
\end{align}
which  is  again  independent  of  $\hat{y}_{t+s}$.  
Therefore,  the  expectation  acts  trivially,  and  summing  over  $s$  yields
\begin{equation}
\log    \frac{\bar{\lambda}_{\delta    \infty}}{\bar{\lambda}_{\infty}}=\mbox{E}[
\log    \frac{\lambda_0    e^{\alpha\delta+\alpha    \hat{y}_t}}
{\lambda_0    e^{\alpha    \hat{y}_t}}
\frac{\lambda_0    e^{\alpha\delta/2^{\gamma}+\alpha    \hat{y}_{t+1}}}
{\lambda_0    e^{\alpha    \hat{y}_{t+1}}}
\frac{\lambda_0    e^{\alpha\delta/3^{\gamma}+\alpha    \hat{y}_{t+2}}}
{\lambda_0    e^{\alpha    \hat{y}_{t+2}}}
\cdots
]
\end{equation}
When    $\gamma>1$,    we        obtain        
\begin{equation}
\log
\frac{\bar{\lambda}_{\delta    \infty}}{\bar{\lambda}_{\infty}}
    <\alpha    \delta(1+\frac{1}{\gamma-1}).
\end{equation}
Then,    the    impact    is    finite.
When    $\gamma=1$,
we        obtain        
\begin{equation}
    \log    \frac{\bar{\lambda}_{\delta    \infty}}{\bar{\lambda}_{\infty}}
    \sim\alpha    \delta\log    T,
\end{equation}
when    $Tgg1$.
We        can    confirm    that    the    impact    is    infinite.
When    $\gamma<1$,
we        obtain        
\begin{equation}
\log        \frac{\bar{\lambda}_{\delta    \infty}}{\bar{\lambda}_{\infty}}
\sim    \alpha    \delta    T^{1-\gamma},
\label{i2}
\end{equation}
when    $T\gg1$.
The    impact    is    infinite.
%%%%%%%%%%%%%%%%%%%%
%Even    with        an    infinite    impact,        the    average        number    of    defaults    converges,    because    the    process    does    not    have    an    absorption        process.
%%%%%%%%%%%%%%%%%%%%%
We  note  that  even  when  the  cumulative  log-impact  diverges
($\gamma  \leq  1$),  the  mean  intensity
$\bar{\lambda}  =  \lambda_0  e^{\alpha^2/2}$
remains  finite  for  all  values  of  $\gamma$,
since  it  depends  on  $\rho_A$  but  not  on  $\gamma$.
The  divergence  manifests  instead  in  the  cumulative
response  to  shocks
$\log(\bar{\lambda}_{\delta\infty}/\bar{\lambda}_\infty)$
and  in  the  anomalous  scaling  of  the  variance
$V(\Lambda_T)$,  as  demonstrated  in  Section~3.

This  contrasts  with  the  Hawkes  and  SE-NBD  processes,
in  which  the  mean  intensity  $\bar{\lambda}$  itself
diverges  in  the  non-steady  phase.
In  the  LGCP  derived  from  the  logistic  Merton  model,
there  is  no  such  absorption  mechanism,
and  the  process  remains  in  a  finite-mean  regime
for  all  $\gamma$.
The  super-normal  transition  therefore  affects
the  estimation  accuracy  of  parameters
(through  the  anomalous  scaling  of  variance)
rather  than  the  stability  of  the  process  itself.

%%%%%%%%%%%%%%%%
It    affects    the    estimation    of        parameters.    
The        Hawkes        process    and    SE-NBD    process    are        examples    of    absorption    processes.
Hence,    there    is    the    absorption    transition,    which    is        a    non-equilibrium    transition.
In    the    next    section    we    consider    the
phase    transition        of    this    process        by        the    direct    calculation    of    the    variance.

In    summary,    the    impact    is    finite    when    $\gamma>1$    and
the    impact    is    infinite    when    $\gamma\leq    1$.    
Here    we    study    the    impact    analysis    after    the    Poisson    limit.
This        is    the    same    phase    transition        which    we    discussed    in        the    discrete    model.    The    discrete    model    is    the    model        before    the    Poisson    limit    \cite{Hisakado8}.
Therefore,        the    limit    inherits    the    properties    of    the    discrete    model.
This    is    one    of    the    reasons    why        this        Poisson    limit    is    important.

    We    identify    a    parallel    structure    between    Hawkes    model    and    this    model    in    the    Poisson    limit.    In    Table    \ref{comp}    we    show    the    comparison    between        the    Hawkes    model    and        the    Merton    model.    Both    processes    have    phase    transitions    before    and    after    the    Poisson    limit.    
    Note  that  we  can  not  compare  the  urn  model  and  correlated  Merton  model  directly  because  of  the  structure  of  the  model,
    However,  we  can  compare  it    after  the  Poisson  limit,  because  both  model  are  Poisson  point  processes.  
Merton    model    has    the    super-normal    phase    transition.    Conversely,    the        Hawkes    and    SE-NBD    models    have    the    transition    between        steady    and    non-steady.    In        the    non-steady    phase,    the    average    of    the    process    diverges.    Therefore,    we    can    confirm    that    the    temporal    correlation    of    self    excitation    is    stronger    than    the    correlation    of        hidden    variables.
In    Appendix    \ref{Aa}.    we        confirm    this    conclusion        using        the    shape    of    the    intensity    functions.

\begin{table}[t]
\centering
\caption{Comparison    between    the    urn    model    approach    and    the        correlated    Merton    model.}
\label{comp}
%\begin{adjustbox}{width=\textwidth}
\begin{tabular}{llll}
\hline
{\footnotesize    Model}    &    {\footnotesize    Discrete    model}    &    {\footnotesize    Poisson    limit}    &    {\footnotesize    Transition}    \\
\hline
{\footnotesize    Urn-model    approach}    &    {\footnotesize        Urn    model    \cite{Hisakado11}}    &    {\footnotesize    Hawkes    process,    SE--NBD}    &    {\footnotesize    Steady    $\leftrightarrow$    Non-steady    }\\
{\footnotesize        Merton    model    }    &    {\footnotesize    Correlated    Merton    model}    &    {\footnotesize    LGCP    with    log-normal    intensity}    &    {\footnotesize    Normal    $\leftrightarrow$    Super-normal}    \\
\hline
\end{tabular}
%\end{adjustbox}
\end{table}

\begin{comment}
\begin{table}[htbp]
        \centering
        \begin{tabular}{|c|c|c|}
        \hline
        Discrete    model    &    urn    model    \cite{Hisakado11}    &    Merton    model    \\
        \hline
        Phase    transition    of    Discrete    model    &    convergence        non-convergence        &    Super-normal        \\
        \hline
        Poisson    limit        &    Hawkes,    SE-NBD    &    LGCP    \\
        \hline
        Phase    transition        of    Poisson    limit    &    steady    non    steady    state        &    Super-normal        \\
        \hline
        \end{tabular}
        \caption{Comparison    of    urn    model    and    Merton    model    before    and    after    Poisson    limit}
        \label{comp}
\end{table}
\end{comment}

%%%%%%%%%%%%%%%%%%%%%%%%%%%%%%%%%%%%%%%%%%%%%%
\section{Super-normal    phase    transition}
\label{3}
The  section  \ref{3}    provides    a    more    rigorous    confirmation    of    the    transition    already    suggested    by    the    impact    analysis.
Here,    we    consider    the        variance    of    $\lambda_T$,    $V(\lambda_T)$.
$\lambda_{T}$    is    the            expected    value        of    the    intensity    function    at        $T$-th    term.
%%%%%%%%%%%%%%

Here,  we  consider  the  variance  of  the  cumulative  intensity  
$\Lambda_T  \equiv  \sum_{t=1}^{T}  \lambda_t$,  
which  we  denote  $V(\Lambda_T)$.
Since  $\mathrm{Var}(\lambda_t)  =  \bar{V}$  for  all  $t$  where    $    \bar{V}=\bar{\lambda}^2    (e^{\alpha^2}-1)$  and  
$\mathrm{Cov}(\lambda_t,  \lambda_{t+i})  =  \bar{V}  d_i$,  we  obtain
\begin{align}
V(\Lambda_T)  =  \bar{V}T  +  2\bar{V}\sum_{i=1}^{T-1}  d_i(T-i).
\label{V1}
\end{align}
The  second  term  arises  from  the  temporal  correlations.
We  examine  the  asymptotic  behavior  of  $V(\Lambda_T)$  as  $T  \to  \infty$.

%%%%%%%%%%%%%%%%%%
\begin{comment}
The    variance    of    $\lambda$    of    the    term    $T$    is    given    by    
\[
\mbox{V}(\lambda_T)=\bar{V}T
+2\bar{V}\sum_{i=1}^{T-1}d_i(T-i),
\]
where    $    \bar{V}=\bar{\lambda}^2    (e^{\alpha^2}-1)$.
The        second    term    is    from    the    temporal    correlation.
We    consider    the    behavior    of    the    second    term
in    limit    $T\to    \infty$.
\end{comment}

\subsection{Exponential    temporal    correlation}
In    this    subsection,    we    study        $V(\Lambda_T)$
for    exponential    decay    model    $d_i=\theta^i,\theta\leq    1$:
\begin{equation}
V(\Lambda_T)\simeq    
\bar{V}T+
2\bar{V}\sum_{i=1}^{T-1}\theta^i    (T-i).    \nonumber
\end{equation}
The    first        term    on    the    right-hand    side    (RHS)    behaves    as    $\propto    T$,    thus,    
this    is    the    normal    diffusion.
In    the    case    that    $\theta    \neq    1$,    the    second        term    is
\[
2\bar{V}[T\frac{1-\theta^{T-1}}{1-\theta}
+\frac{(T-1)\theta^{T-1}(1-\theta)-(1-\theta^{T-1})\theta}{(1-\theta)^2}]        
\propto    T
\]
and        normal    diffusion.
We    conclude    that    as    the    number    of    data    samples    increases,    
the    variance    increases    as    $T$.
When    $\theta=1$,    there    is    no    temporal    correlation    decay        and    all    obligors    are    correlated    $\rho_A$.    Hence,    there    is    no    phase    transition    for    $\theta<1$.

%%%%%%%%%%%%%

\begin{figure}[htbp]
        \centering
        \includegraphics[width=0.9\textwidth]{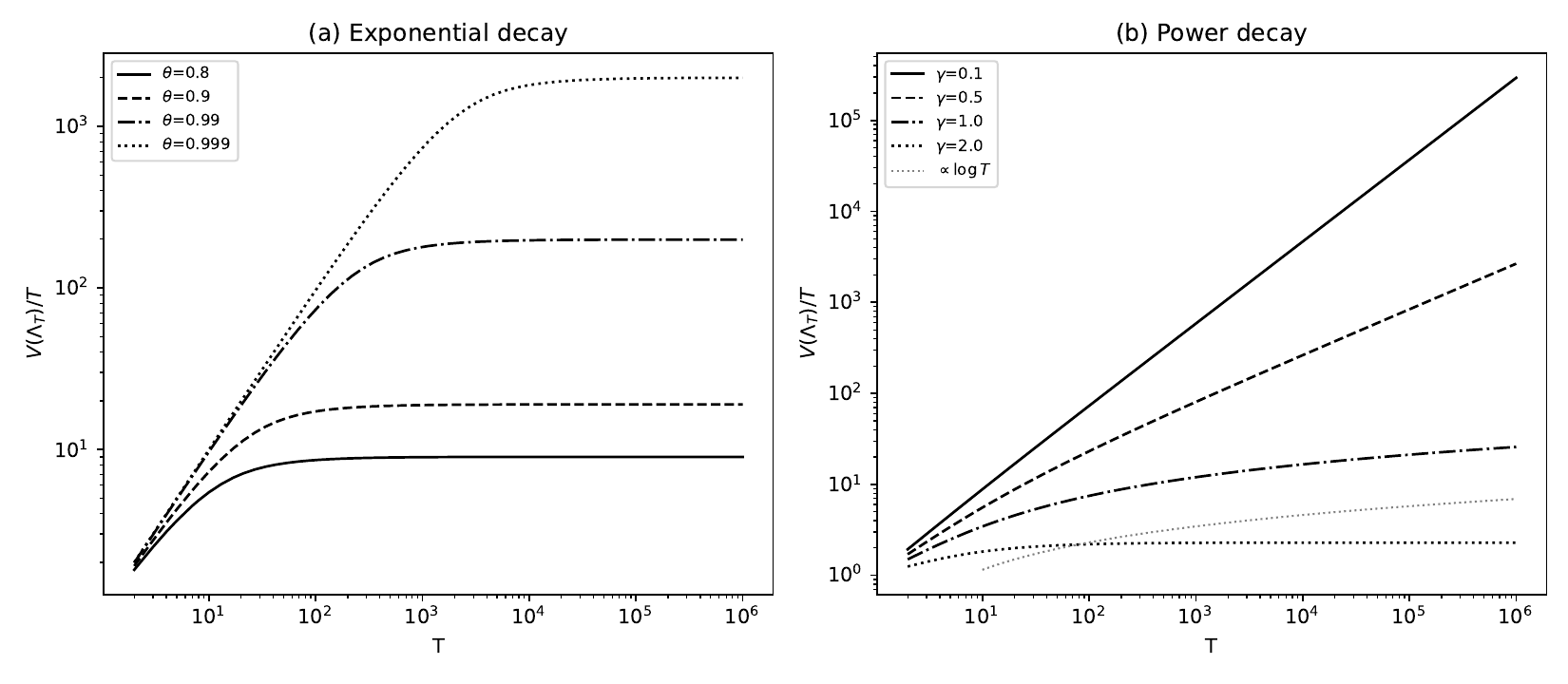}
        \caption{
Double-logarithmic  plot  of  the  normalized  cumulative  variance
$V(\Lambda_T)/T$  as  a  function  of  $T$,  where  $\Lambda_T  =  \sum_{t=1}^{T}\lambda_t$.
Since  $V(\Lambda_T)\sim  T$  in  the  normal-diffusion  regime,  this  normalization
isolates  the  deviation  from  normal  diffusion:  $V(\Lambda_T)/T$  saturates  to  a
constant  when  the  process  is  normally  diffusive,  while  it  diverges  as
$T\to\infty$  whenever  the  diffusion  is  anomalous.
(a)  Exponential  decay  model  with  $\theta  \in  \{0.8,  0.9,  0.99,  0.999\}$:  for
all  $\theta<1$,  $V(\Lambda_T)/T$  saturates  to  the  finite  constant
$1+2\theta/(1-\theta)$,  confirming  normal  diffusion  and  the  absence  of  a
phase  transition  for  $\theta<1$.
(b)  Power  decay  model  with  $\gamma  \in  \{0.1,  0.5,  1.0,  2.0\}$:  for
$\gamma>1$,  $V(\Lambda_T)/T$  likewise  saturates  to  a  constant  (normal
diffusion);  at  the  critical  point  $\gamma=1$  it  grows  as  $\log  T$  (gray
dotted  reference  line);  and  for  $\gamma<1$  it  diverges  as  a  power  law,
$T^{1-\gamma}$.  The  qualitative  contrast  between  a  saturating  curve
($\gamma>1$)  and  a  diverging  curve  ($\gamma\le1$)  provides  a  clear  visual
signature  of  the  super-normal  phase  transition  at  $\gamma=1$
}
        \label{impact2}
\end{figure}

Thus,  we  calculate  $V(\Lambda_T)/T$  numerically.  We  set  $\theta  \in  \{0.8,  0.9,  0.99,  0.999\}$.  Fig.~\ref{impact2}  (a)  shows  the  double-logarithmic  plot  of  $V(\Lambda_T)/T$.  Here,  it  is  clearly  seen  that  $V(\Lambda_T)/T$  saturates  to  a  finite  constant  as  $T  \to  \infty$  for  every  $\theta<1$,  confirming  that  the  process  is  normally  diffusive  and  that  there  is  no  phase  transition  for  $\theta  <  1$.

\begin{comment}
\begin{figure}[h]
\begin{center}
\begin{tabular}{c}
%    1
\begin{minipage}{0.5\hsize}
\begin{center}
\includegraphics[clip,    width=7cm]{2429.eps}
\hspace{1.6cm}    (a)
\end{center}
\end{minipage}
%    2
\begin{minipage}{0.5\hsize}
\begin{center}
\includegraphics[clip,    width=7cm]{2428.eps}
    \hspace{1.6cm}    (b)
\end{center}
\end{minipage}
%    3
    \end{tabular}
\caption{Plot    of    $V    (\lambda_T)/V(\lambda(y))$        (a)    exponential    decay    and    (b)    power    decay.    We    calculate        $\theta\in    \{0.8,    0.9,0.99,    0.999\}$    for    (a)    and        $\gamma\in    \{0.1,    0.5,1.0,    2.0\}$    for    (b).    
In    (a),        $V    (\lambda_T)/V(\lambda(y))$    converges    as        a    normal    distribution.    
In    (b),    it    converges    slower    than    the    normal    distribution    when    $\gamma\leq    1$    and    converges    as    the    normal    distribution            when    $\gamma>1$.    
}
\label{impact2}
\end{center}
\end{figure}
\end{comment}

\subsection{Power    temporal    correlation}

In    this    subsection,    we    consider
the    power    decay    case    $d_i=1/(i+1)^{\gamma},i=1,2,\cdots$,
where    $\gamma\ge    0$    is    the    power    index.
The    power    correlation    affects    the    number    of    defaults    over    long    periods    of    time.
Ranges    $\gamma\leq1$    and    $\gamma>1$    are    called    long    memory    and    intermediate    memory,    respectively    \cite{Long}.
In    contrast,    the    exponential    decay    affects    short    periods    of    time    and    is    called    short    memory.
The    asymptotic    behavior    of    $V(\Lambda_T))$    is    given    as:
\[
\mbox{V}(\Lambda_T)\simeq    \bar{V}T
+2\bar{V}T\sum_{i=1}^{T-1}(i+1)^{-\gamma}(T-i).    
\]

\subsubsection{$\gamma>1$    case}
    We        obtain
    \begin{eqnarray}
        \mbox{V}(\Lambda_T)    
        &\simeq    &\bar{V}T
        +2\bar{V}\sum_{i=1}^{T-1}    (T-i)/(i+1)^{\gamma}\nonumber    \\
        &\simeq    &    
        \bar{V}T
        +2\bar{V}T^{-\gamma+2}/(\gamma-1).
        \label{g1}
\end{eqnarray}    
    The    first    term        is    the    normal    diffusion
    and    the    second        term        is    proportional    to    the        $T^{-\gamma+2}$,    where    $\gamma>1$.
    It    is        slower        than        normal    diffusion.
    Therefore,    the    significant    terms    are    the    first        term    which    is    the    normal    diffusion.
\subsection{$\gamma=1$  case}
$\mbox{V}(\Lambda_T)$    behaves    as
\begin{equation}
\mbox{V}(\Lambda_T)\simeq        \bar{V}T
+2\bar{V}\sum_{i=1}^{T-1}    (T-i)/(i+1).
\label{14}
\end{equation}  
The    RHS    of    Eq.    (\ref{14})    is    evaluated    as    
\begin{equation}
        \mbox{V}(\Lambda_T)
        \simeq        
        \bar{V}T+
        +2\bar{V}T\log    T-T+2]\sim    T\log    T.
\end{equation}
    In    conclusion,    $\mbox{V}(\Lambda_T)$    behaves    asymptotically    as
    \begin{equation}
\mbox{V}(\Lambda_T)    \sim    T\log    T
    \end{equation}
    and        becomes    an        anomalous        faster    diffusion.

    \subsubsection{$\gamma<1$    case}
    
    $\mbox{V}(\Lambda_T)$    is    calculated    as
    \begin{equation}
        \mbox{V}(\Lambda_T)
        \simeq        
        \bar{V}T
        +2\bar{V}
        [\frac{1}{(1-\gamma)(2-\gamma)]}T^{-\gamma+2}\sim    T^{-\gamma+2}.
\end{equation}    
Thus,    we    can    conclude    that    $\mbox{V}(\Lambda_T)$    behaves    as
    \begin{equation}
    \mbox{V}(\Lambda_T)    \sim    T^{-\gamma+2},
    \label{g2}
\end{equation}
which    is        also    the    anomalous        faster        diffusion.

In    conclusion,    a    phase    transition    occurs
when    the    temporal    correlation    decays    by    the    power    law.
It    is    the    same    as    the    original    process    which    is    before    the    limit.
When    the    power    index,    $\gamma$,    is    less    than    one,    the    variance,    $\mbox{V}(\Lambda_T)$,    exhibits
anomalous    diffusion.    
Conversely,    when    the    power    index,    $\gamma$,    is    greater    than    one,
it    is    the    normal    diffusion.
This    phase    transition,    referred        to    as    a    super-normal    transition    \cite{hod,Hisakado2},
marks    the    transition    between    long    memory    and    intermediate    memory.
In    that    study,    when    the    power    index    was    less    than    one,    the    estimator
does        not    converge            to    the    steady    state    \cite{Hisakado6}.

%%%%%%%%%%%%%%%%%%%%%

To  confirm  the  phase  transition,  we  calculate  $V(\Lambda_T)/T$.  Fig.~\ref{impact2}  (b)  shows  the  double-logarithmic  plot  of  $V(\Lambda_T)/T$  for  $\gamma  \in  \{0.1,  0.5,  1.0,  2.0\}$.  For  $\gamma>1$,  $V(\Lambda_T)/T$  saturates  to  a  finite  constant,  indicating  normal  diffusion.  At  the  critical  point  $\gamma=1$,  $V(\Lambda_T)/T$  diverges  logarithmically.
For  $\gamma<1$,  $V(\Lambda_T)/T$  diverges  as  a  power  law.
This  qualitative  distinction  between  a  saturating  curve  ($\gamma>1$)  and  a  diverging  curve  ($\gamma  \leq  1$)  provides  direct  numerical  confirmation  of  the  super-normal  phase  transition  at  $\gamma=1$.

\section{Estimation    of    parameters}
\label{4}

Since    the    existence    of    the    super-normal    transition    depends    on    whether    $\gamma$    is    below    or    above    1,    empirical    estimation    of    $\gamma$    is    crucial.    We    therefore    estimate    $\gamma$    from    historical    default    data    and    examine    whether    real    portfolios    operate    in    the    normal    or    super-normal    regime.

The    temporal    correlation    of    macroeconomic    conditions    plays    a    crucial    role    in    modeling    default    events    using    a    Merton-type    Poisson    framework.    In    this    model,    the    number    of    defaults    in    each    year    is    assumed    to    follow    a    Poisson    distribution    with    intensity    $\lambda_t    =    \lambda_0    e^{\alpha    y_t}$,    where    $y_t$    is    a    latent    standard    normal    variable    representing    the    macroeconomic    environment.    The    time    series    $\{y_t\}$    exhibits    autocorrelation,    which    reflects    the    persistence    of    economic    conditions.    The    functional    form    of    this    temporal    correlation—either    exponential    decay    with    parameter    $\theta$    or    power        decay    with    exponent    $\gamma$—can    produce    a    super-normal    phase    transition.    This    transition    affects    the    behavior    of    the    default    distribution    and    the    convergence    properties    of    parameter    estimates.

In    this    section,    we    estimate    the    following    key    model    parameters:    the    baseline    intensity    $\lambda_0$,        macro-sensitivity    coefficient    $\alpha$,    and        temporal    correlation    parameter    $\theta$    or    $\gamma$.    The    estimation    is    performed    using    annual    default    data    from    Moody’s    (1920–2023)    and    S\&P    (1981–2023),    across    three    rating    categories:    all    firms    (ALL),        speculative    grade    (SG),    and    investment    grade    (IG)    \cite{Data1,    Data2}.    We    adopt    a    Bayesian    inference    approach    implemented    in    Stan,    and    report    the    maximum    a    posteriori    (MAP)    estimates.    The    prior    distributions    for    the    parameters    are    constructed    based    on    preliminary    maximum    likelihood    estimates    and    autocorrelation    analysis    of    the    inferred    latent    factors.

To    ensure    comparability    across    years    and    rating    categories,    we    normalize    the    number    of    defaults    by    rescaling    them    to    a    common    portfolio    size    of    3000    obligors    per    year.    Specifically,    we    compute    the    default    count    in    year    $t$    as    $k_t^{*}    =    (k_t    /    n_t)    \times    3000$,    where    $k_t$    is    the    observed    number    of    defaults    and    $n_t$    is    the    number    of    obligors    in    that    year.    This    normalization    is    essential    not    only    for    visual    comparison    but    also    for    statistical    analysis;    without    it,    the    time    series    would    be    heteroscedastic    due    to    varying    portfolio    sizes,    undermining    the    assumptions    of    stationarity    required    by    tests    such    as    the    augmented    Dickey–Fuller    (ADF)    test.

\begin{figure}[htbp]
        \centering
        \includegraphics[width=0.9\textwidth]{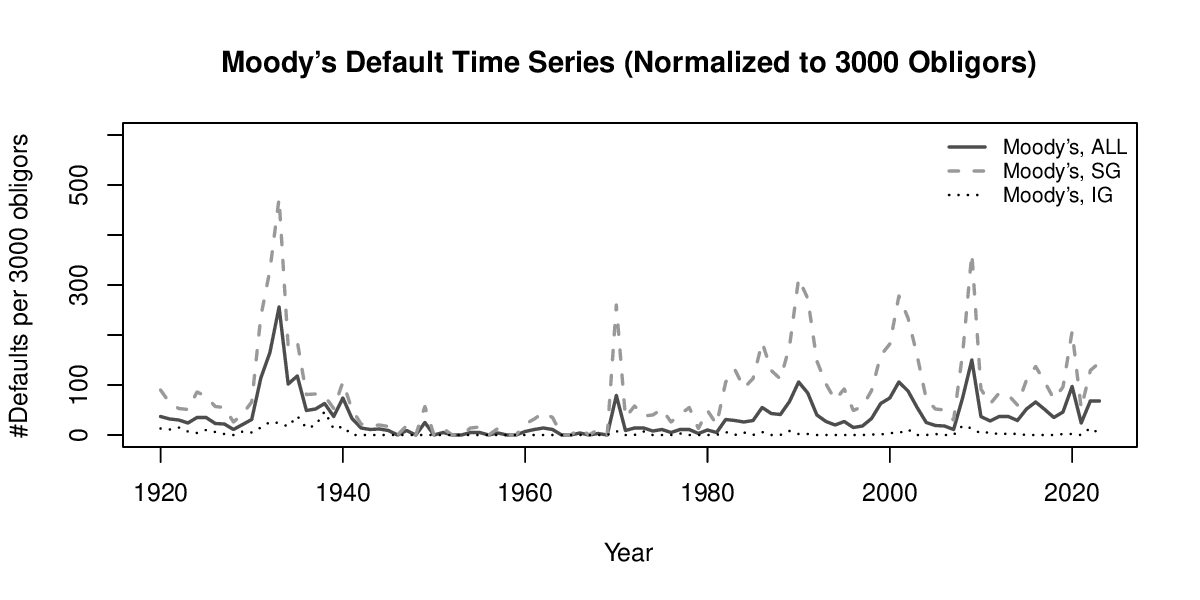}
        \caption{Annual    default    counts    per    3000    obligors    for    Moody’s-rated    firms    from    1920    to    2023.}
        \label{fig:moody-defaults}
\end{figure}

Figure~\ref{fig:moody-defaults}    displays    the    time    series    of    annual    defaults    per    3000    obligors    for    Moody’s-rated    firms    from    1920    to    2023.    We    plot    separate    curves    for    all    ratings    (ALL),    speculative    grade    (SG),    and    investment    grade    (IG)    categories.    The    visualized    trends    show    notable    peaks    during    economic    downturns,    particularly    in    the    SG    category,    while    the    IG    defaults    remain    relatively    rare    and    stable    throughout    the    period.

\subsection{A.    Preliminary    Estimation    Using    Maximum    Likelihood}

As    a    preliminary    step,    we    estimate    the    parameters    $\lambda_0$    and    $\alpha$    by    maximizing    the    likelihood    function    under    the    assumption    that    the    latent    macroeconomic    factor    $y_t$    is    temporally    independent.    Using    the    resulting    estimates,    we    then    infer    the    value    of    $y_t$    for    each    year    via    the    transformation
\[
y_t    =    \frac{\log(k_t    +    0.5)    -    \log(\lambda_0)}{\alpha},
\]
where    $k_t$    is    the    normalized    default    count    per    3000    obligors.    This    transformation    assumes
that    $k_t\simeq    E[N_t|y_t]=\lambda_t    =    \lambda_0    e^{\alpha    y_t}$    holds    pointwise.

Subsequently,    we    compute    the    autocorrelation    function    (ACF)    of    the    inferred    $\{y_t\}$    series    to    assess    the    presence    and    nature    of    temporal    dependence.    The    shape    of    the    ACF    provides    empirical    guidance    in    distinguishing    between    exponential    and    power        decay    patterns.    These    findings    form    the    basis    for    constructing    the    prior    distributions    of    temporal    correlation    parameters    in    the    Bayesian    estimation    framework.

\begin{table}[htbp]
\centering
\caption{Maximum    likelihood    estimates    (MLE)    of    $\lambda_0$    and    $\alpha$,    and    empirical    autocorrelation    values    $\theta$    and    $\gamma$.    Standard    errors    are    in    parentheses.}
\label{tab:mle-acf}
\begin{tabular}{lcccc}
        \hline        
Dataset    &    MLE    $\lambda_0$    &    MLE    $\alpha$    &    ACF    $\theta$    &    ACF    $\gamma$    \\
    \hline    
%        \midrule
Moody’s    ALL    (1920–2023)        &    18.1    (0.1)    &    1.4    (2.6)    &    0.890    (0.004)    &    0.64    (0.09)    \\
Moody’s    SG    (1920–2023)        &    42.4    (0.1)    &    1.6    (6.7)    &    0.880    (0.005)    &    0.63    (0.09)    \\
Moody’s    IG    (1920–2023)        &    0.6    (0.1)        &    2.5    (0.4)    &    0.84    (0.01)        &    0.74    (0.07)    \\
Moody’s    ALL    (1980–2023)        &    39.9    (3.9)    &    0.61    (0.07)    &    0.75    (0.11)    &    1.0    (0.7)    \\
Moody’s    SG    (1980–2023)        &    107.3    (9.7)    &    0.58    (0.07)    &    0.72    (0.02)    &    1.24    (0.09)    \\
Moody’s    IG    (1980–2023)        &    1.0    (0.2)    &    1.6    (0.4)    &    0.70    (0.08)    &    0.95    (0.52)    \\
S\&P    ALL    (1981–2023)        &    36.8    (0.1)    &    0.62    (3.63)    &    0.74    (0.12)    &    1.01    (0.73)    \\
S\&P    SG    (1981–2023)        &    100.2    (9.6)    &    0.62    (0.07)    &    0.74    (0.05)    &    0.86    (0.39)    \\
S\&P    IG    (1981–2023)        &    1.0    (0.2)    &    1.5    (0.4)    &    0.68    (0.06)    &    0.98    (0.55)    \\
%\bottomrule
    \hline    
\end{tabular}
\end{table}

Table~\ref{tab:mle-acf}    presents    the    preliminary    estimates.
The    MLE    results,    based    on    the    assumption    of    no    temporal    correlation,    provide    baseline    estimates    for    the    intensity    parameter    $\lambda_0$    and    the    macroeconomic    sensitivity    $\alpha$.    The    ACF-based    values    of    $\theta$    and    $\gamma$,    derived    from    the    inferred    latent    factor    $y_t$,    offer    preliminary    insight    into    the    underlying    temporal    structure.    Notably,    many    of    the    $\gamma$    values    fall    below    1,    indicating    the    potential    presence    of    long-range    dependence.

\begin{figure}[htbp]
        \centering
        \includegraphics[width=0.9\textwidth]{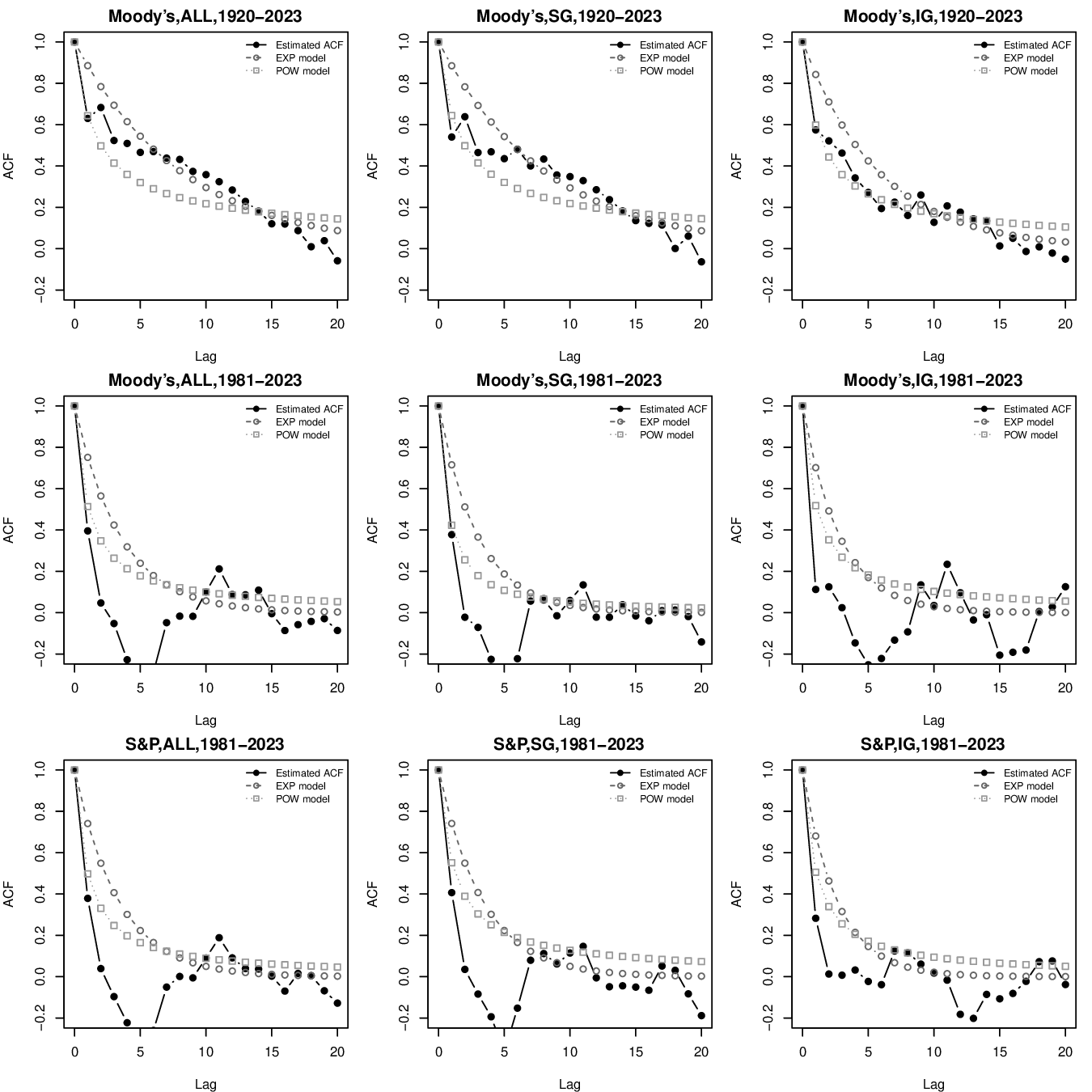}
        \caption{
        Sample    autocorrelation    functions    (ACF)    of    the    inferred    latent    macroeconomic    factor    $y_t$    for    nine    rating    segments,    with    fitted    exponential    and    power        decay    models.    Each    panel    corresponds    to    a    different    data    subset    (Moody’s    or    S\&P,    across    All,    Speculative    Grade,    and    Investment    Grade).    The    solid    black    points    represent    the    empirical    ACF    values.    The    dashed    gray    lines    with    circles    show    exponential    decay    fits,    and        the    dotted    gray    lines    with    boxes    show    power        decay    fits.    The    shape    of    the    ACF    helps    determine    whether    short-    or    long-range    dependence    is    present    in    each    case.
        }
        \label{fig:acf-fit-grid}
\end{figure}

Figure~\ref{fig:acf-fit-grid}    displays    the    sample    autocorrelation    function    (ACF)    of    the    inferred    latent    variable    $y_t$,    alongside    fitted    curves    assuming    exponential    and    power        decay.    We    discuss    two    illustrative    cases    in    detail    below.

For    the    Moody’s    all-rated    dataset    over    the    full    period    (1920–2023).  
The  data  is  annual  data  and  the  number  of  data  is    104.
For    the  S\&P    all-rated    dataset    over    the    period    (1981–2023).
The  data  is  annual  data  and  the  number  of  data  is    43.
The    maximum    likelihood    estimates    of    the    intensity    parameters    are    $\hat{\lambda}_0    =    18.1\    (0.1)$    and    $\hat{\alpha}    =    1.4\    (2.6)$.    
The    estimated    decay    parameter    from    exponential    fitting    is    $\hat{\theta}    =    0.89\    (0.004)$,    while    the    power        model    yields    $\hat{\gamma}    =    0.64\    (0.09)$.    The    relatively    low    value    of    $\gamma$    indicates    the    presence    of    long-range    dependence    in    the    default    process.

In    contrast,    the    results    for    the    more    recent    period    (1980–2023)    show    a    different    pattern.    The    estimates    are    $\hat{\lambda}_0    =    39.9\    (3.9)$    and    $\hat{\alpha}    =    0.61\    (0.07)$,    with    $\hat{\theta}    =    0.75\    (0.11)$    and    $\hat{\gamma}    =    1.0\    (0.7)$.    The    higher    value    of    $\gamma$,    approaching    or    exceeding    the    critical    threshold    of    1,    suggests        that    long-range    dependence    is    no    longer    evident    in    the    modern    credit    environment.    This    comparison    indicates    a    possible    structural    transition    in    the    nature    of    temporal    correlation    in    default    events    over    time.

\subsection{B.    Bayesian    Parameter    Estimation}

We    now    present    the    results    of    Bayesian    parameter    estimation    for    the    Merton-Poisson    model    under    both    exponential    and    power        decay    assumptions.    The    prior    distributions    were    constructed    based    on    the    maximum    likelihood    estimates    and    ACF    analysis    described    in    the    previous    subsection.    Specifically,    normal    priors    were    placed    on    each    parameter,    centered    at    the    MLE    values    with    standard    deviations    scaled    by    hyperparameter    $sc$    times    the    standard    error.    This    approach    reflects    prior    knowledge    while    allowing    sufficient    flexibility    for    posterior    updating.

The    estimation    was    implemented    in    Stan    (version    2.19.2)    via    R    (version    3.6.2),    and    the    maximum    a    posteriori    (MAP)    estimates    were    used    for    comparison.    The    details    are    provided    in    Appendix    \ref{Ab}.

%\begin{comment}
\begin{table}[htbp]
\centering
\caption{
Bayesian    parameter    estimates    for    each    dataset    under    the    exponential    and    power        temporal    correlation    models.    MAP    estimates    are    reported    in    parentheses    with    standard    errors.}
\label{tab:parameter-estimates}
%\resizebox{18cm}{!}{        %    横幅を18cmにスケーリング
%\begin{adjustbox}{width=\textwidth}
\begin{tabular}{lcccccc}
    %    \toprule
        \hline    
Dataset    
&    $\lambda_0$    (Exp)    &    $\alpha$    (Exp)    &    $\theta$    
&    $\lambda_0$    (Pow)    &    $\alpha$    (Pow)    &    $\gamma$    \\
    \hline    
%        \midrule
Moody’s    ALL    (1920–2023)        &    18.1    (0.4)    &    1.6    (0.2)    &    0.88    (0.01)    &    18.1    (0.4)    &    1.3    (0.3)    &    0.4    (0.2)    \\
Moody’s    SG    (1920–2023)        &    42.5    (1.4)    &    1.9    (0.3)    &    0.83    (0.04)    &    42.4    (0.3)    &    1.4    (0.2)    &    0.6    (0.1)    \\
Moody’s    IG    (1920–2023)        &    0.6    (0.1)        &    2.7    (0.2)    &    0.84    (0.01)    &    0.6    (0.1)        &    2.4    (0.2)    &    0.72    (0.06)    \\
Moody’s    ALL    (1980–2023)        &    39.3    (5.4)    &    0.65    (0.08)    &    0.59    (0.11)    &    39.8    (3.5)    &    0.63    (0.06)    &    1.1    (0.4)    \\
Moody’s    SG    (1980–2023)        &    107.0    (9.1)    &    0.66    (0.05)    &    0.70    (0.02)    &    107.4    (8.7)    &    0.59    (0.05)    &    1.25    (0.09)    \\
Moody’s    IG    (1980–2023)        &    1.0    (0.2)    &    1.8    (0.3)    &    0.62    (0.07)    &    0.95    (0.2)    &    1.7    (0.3)    &    1.4    (0.4)    \\
S\&P    ALL    (1981–2023)        &    36.8    (0.2)    &    0.64    (0.22)    &    0.52    (0.15)    &    36.8    (0.1)    &    0.67    (0.16)    &    1.1    (0.5)    \\
S\&P    SG    (1981–2023)        &    100.1    (8.9)    &    0.69    (0.05)    &    0.68    (0.04)    &    100.1    (8.9)    &    0.64    (0.05)    &    1.1    (0.3)    \\
S\&P    IG    (1981–2023)        &    1.0    (0.2)    &    1.6    (0.3)    &    0.65    (0.06)    &    1.0    (0.2)    &    1.6    (0.3)    &    1.2    (0.4)    \\
%    \bottomrule
    \hline    
\end{tabular}
%\end{adjustbox}
%}
\end{table}
%\end{comment}

Table~\ref{tab:parameter-estimates}    summarizes    the    estimated    parameters    across    all    datasets    using    the    Bayesian    framework.    The    Bayesian    estimates    are    reported    separately    for    exponential    and    power    correlation    models.    Overall,    the    MAP    estimates    tend    to    be    close    to    their    MLE    counterparts,    with    narrower    standard    errors,    reflecting    the    informative    nature    of    the    priors    derived    from    preliminary    analysis.    The    estimated    decay    parameters    $\theta$    and    $\gamma$    in    the    Bayesian    models    generally    align    with    those    obtained    from    the    ACF    fitting,    confirming    the    robustness    of    the    temporal    correlation    structure    inferred    from    the    data.    For    several    datasets,    especially    those    covering    longer    time    periods    such    as    Moody’s    ALL    (1920–2023),    the    estimated    $\gamma$    is    significantly    less    than    1,    further    supporting    the    existence    of    persistent    temporal    correlation    in    macroeconomic    risk    factors.

To    assess    the    relative    performance    of    the    exponential    and    power    temporal    correlation    models,    we    evaluate    three    model    selection    criteria:    the    Widely    Applicable    Information    Criterion    (WAIC),    the    Widely    Applicable    Bayesian    Information    Criterion    (WBIC),    and    the    Leave-Future-Out    cross-validation    score    (LFO).    WAIC    is    a    fully    Bayesian    measure    of    out-of-sample    predictive    accuracy    based    on    the    log    pointwise    predictive    density,    and    is    robust    in    the    presence    of    overfitting.    WBIC    approximates    the    marginal    likelihood    by    evaluating    the    expectation    of    the    log-likelihood    under    a    tempered    posterior    distribution.    Unlike    the    classical    BIC,    WBIC    is    valid    in        both    regular    and    singular    statistical    models.    The    LFO    score,    a    time-series    analogue    of    leave-one-out    cross-validation,    evaluates    the    model’s    ability    to    predict    future    data    by        holding    out    the    most    recent    observation    at    each    time    step.    Lower    values    of    WAIC,    WBIC,    and    LFO    indicate    better    generalization    performance.

%\begin{comment}
\begin{table}[htbp]
\centering
\caption{
    The        model    comparison    results        are    based    on    three    information    criteria:    WAIC,    WBIC    (Widely    Applicable    Bayesian    Information    Criterion),    and    LFO    (Leave-Future-Out    cross-validation).    Scores    are    reported    for    both        the    exponential    and    power    decay    models    for    each    dataset.    Lower    values    indicate    better    model    fit.    The    best    score    for    each    criterion    is    indicated        in    bold.}
\label{tab:model-comparison}
%\resizebox{18cm}{!}{        %    横幅を18cmにスケーリング
%\begin{adjustbox}{width=\textwidth}
\begin{tabular}{lrrrrrr}
        \toprule
Dataset    &    LFO    (Exp)    &    WAIC    (Exp)    &    WBIC    (Exp)    &    LFO    (Pow)    &    WAIC    (Pow)    &    WBIC    (Pow)    \\
%        \midrule
\hline
Moody's    ALL    (1920–2023)    &    58.42    &    700.22    &    953.28    &    \textbf{56.99}    &    \textbf{683.99}    &    \textbf{943.76}    \\
Moody's    SG    (1920–2023)    &    70.78    &    804.80    &    1151.05    &    \textbf{70.54}    &    \textbf{791.15}    &    \textbf{1137.41}    \\
Moody's    IG    (1920–2023)    &    24.68    &    308.80    &    511.31    &    \textbf{23.45}    &    \textbf{296.92}    &    \textbf{489.06}    \\
Moody's    ALL    (1980–2023)    &    59.28    &    \textbf{310.93}    &    392.02    &    \textbf{58.86}    &    311.64    &    \textbf{390.80}    \\
Moody's    SG    (1980–2023)    &    68.64    &    353.92    &    442.69    &    \textbf{67.99}    &    \textbf{353.19}    &    \textbf{440.90}    \\
Moody's    IG    (1980–2023)    &    27.62    &    145.26    &    203.29    &    \textbf{27.53}    &    \textbf{140.53}    &    \textbf{200.93}    \\
S\&P    ALL    (1981–2023)    &    56.58    &    \textbf{307.59}    &    \textbf{387.70}    &    \textbf{56.30}    &    307.63    &    388.95    \\
S\&P    SG    (1981–2023)    &    66.55    &    \textbf{350.14}    &    437.62    &    \textbf{66.31}    &    350.95    &    \textbf{435.15}    \\
S\&P    IG    (1981–2023)    &    \textbf{14.57}    &    146.71    &    \textbf{198.38}    &    15.32    &    \textbf{142.43}    &    202.04    \\
%    \bottomrule
\hline
\end{tabular}
%\end{adjustbox}
%}    
%    end    resizebox
\end{table}
%\end{comment}

Table~\ref{tab:model-comparison}    presents    the    model    comparison    results    based    on    LFO,    WAIC,        and    WBIC.    
For    each    dataset,    the    prior    scale    hyperparameter    $sc$    was    optimized    to    minimize    the    LFO    score,    ensuring    comparability    between    models    with    different    levels    of    regularization.

The    results    consistently    demonstrate    that    the    power        decay    model    outperforms    the    exponential    decay    model    in    many    cases,    especially    for    datasets    spanning    longer    historical    periods    such    as    Moody’s    ALL,    SG,    and    IG    (1920–2023).    In    these    cases,    LFO    and    WAIC    values    are    substantially    lower    for    the    power    model,    indicating    superior    predictive    performance.    Importantly,    the    WBIC    values    also    favor    the    power    specification,    showing        that    the    marginal    likelihood    is    higher    for    models    incorporating    long-range    dependence.

For    more    recent    and    shorter    datasets—such    as    Moody’s    IG    (1980–2023)    and    S\&P    IG    (1981–2023)—the    differences    between    the    two    models    are        less    pronounced.    Although    the    exponential    model    yields    slightly    better    LFO    scores    in    some    cases,    the    WAIC    and    WBIC    values    remain    close,    indicating    comparable    adequacy    of    both    models.    These    patterns    may    reflect    a    structural    shift    in    the    temporal    dynamics    of    credit    risk,    with    long-range    dependence    becoming    less    significant    in    the    modern    era.

Overall,    the    alignment    of    all    three    metrics—LFO,    WAIC,    and    WBIC—lends    strong    support    to    the    power    model    for    long-horizon    data,    while    endorsing    the    exponential    model    as    a    parsimonious    alternative    for    shorter    and    more    recent    time    series.

The    empirical    estimates    indicate    that    pre-1980    portfolios    were    often    in    or    near    the    super-normal    regime    ($\gamma\leq    1$),    whereas    more    recent    portfolios    are    typically    closer    to    the    normal    regime    ($\gamma>1$).    Thus,    the    theoretical    phase    transition    discussed    above    is    not    merely    a    mathematical    artifact    but    appears    to    have    practical    relevance    for    historical    credit    markets.
%%%%%%%%%%%%%%%%%%%%%
The  theoretical  results  in  Section~\ref{3}  predict  that,
under  power-law  temporal  decay  with  exponent  $\gamma$,
the  variance  of  the  cumulative  default  count  $\Lambda_T$
scales  as  Eq.(\ref{scale}).
In  principle,  this  scaling  could  be  examined  empirically
by  computing  $V(\Lambda_T)$  for  increasing  aggregation
windows  $T$  and  testing  whether  the  slope  in  a
double-logarithmic  plot  deviates  from  unity.
Such  an  analysis  would  provide  a  more  direct  empirical
counterpart  of  the  theoretical  phase  transition.

However,  the  available  time  series  are  too  short
for  a  reliable  discrimination  of  these  scaling  behaviors.
For  the  Moody's  dataset  (1920--2023,  $T  =  104$  years),
distinguishing  $T^{2-\gamma}$  from  $T$  would  require
the  slope  to  differ  detectably  from  unity  over  the
observed  range  of  $T$,  which  is  not  feasible  given
the  limited  sample  size.
For  example,  with  $\gamma  =  0.6$  the  predicted  scaling
is  $T^{1.4}$,  but  over  a  range  of  $T  \leq  104$
this  is  difficult  to  distinguish  from  $T$  in  practice.
We  leave  a  direct  empirical  scaling  analysis
to  future  work.
%%%%%%%%%%%%%%%%%%%%
\section{Summary  of  Main  Results}
\label{5}
In  this  section,  we  summarize  the  principal  theoretical  results  
of  this  paper  in  the  form  of  propositions.

\begin{proposition}[LGCP  limit  of    generalized  Merton  model]
Consider  the  correlated  Merton  model  with  logistic  CDF,  
in  which  the  latent  Gaussian  variables  $\{y_t\}$  have  
temporal  correlation  matrix  $\Sigma$  with  entries  $d_i$.  
Under  the  double  scaling  limit  $p'  \to  0$,  $N  \to  \infty$  
with  $\lambda_0  =  Np'^{1/\sqrt{1-\rho_A}}$  fixed,  
the  default  count  process  $\{N_t\}$  converges  to  a  
Log-Gaussian  Cox  Process  (LGCP)  with  log-normal  intensity
\begin{align}
\lambda(\hat{y})  =  \lambda_0  e^{\alpha  \hat{y}},  
\quad  \alpha  =  \frac{\sqrt{\rho_A}}{\sqrt{1-\rho_A}}\beta,
\end{align}
where  the  intensity  distribution  $f(\lambda)$  is  log-normal  
with  mean  $\bar{\lambda}  =  \lambda_0  e^{\alpha^2/2}$  
and  variance  $\bar{V}  =  \bar{\lambda}^2(e^{\alpha^2}-1)$.
The  temporal  correlation  structure  of  $\{\hat{y}_t\}$  
is  inherited  from  the  original  model  via  $\{d_i\}$.
\end{proposition}
%%%%%%%%%%%%%%%%%
\begin{proof}
The  result  follows  from  Eqs.~(\ref{G})--(\ref{f}),  
together  with  the  factorization  of  $G(y)$  
in  Eq.~(\ref{G})  and  the  change  of  variables  
from  $\hat{y}$  to  $\lambda$  in  Eq.~(\ref{f}).
\end{proof}
%%%%%%%%%%%%%%%%%%%%%%
\begin{remark}
This  proposition  identifies  the  precise  role  of  the  Merton  
model  structure  in  the  derivation.
The  factor  decomposition  in  Eq.(\ref{G})  separates  the  
baseline  default  probability  $p'$  from  the  macroeconomic  
factor  $y_t$,  and  dictates  the  double  scaling  limit  
whose  exponent  $1/\sqrt{1-\rho_A}$  originates  from  
the  asset  correlation  $\rho_A$.
Under  the  standard  Poisson  limit  $\lambda_0  =  Np$,  
the  dependence  on  $y_t$  vanishes  and  all  temporal  
correlations  are  lost;  the  Merton-specific  scaling  
is  therefore  the  mechanism  by  which  temporal  correlations  
survive  the  Poisson  limit.
The  LGCP  limit  itself  requires  not  the  Merton  model  structure  
specifically,  but  that  the  asset  CDF  has  an  exponential  
tail  $\Phi(x)  \sim  Ae^{\kappa  x}$  as  $x  \to  -\infty$(see  Appendix~\ref{Ac}  and  Table~\ref{tab:tail}).
This  condition  is  satisfied  by  the  logistic  and  
generalized  hyperbolic  distributions,  but  not  by  
the  Gaussian  or  $t$-distributions.
The  connection  between  the  Merton  model  and  the  LGCP  
via  this  double  scaling  limit  appears  to  be  new  
in  the  literature.
\end{remark}
%%%%%%%%%%%%%
\begin{proposition}[Super-normal  phase  transition]
\label{thm:phase}
Let  $\Lambda_T  =  \sum_{t=1}^{T}\lambda_t$  be  the  
cumulative  intensity  of  the  LGCP  in  Proposition~1.  
The  asymptotic  behavior  of  $V(\Lambda_T)$  as  $T  \to  \infty$  
is  determined  by  the  temporal  correlation  structure  as  follows.
\begin{enumerate}
\item[(i)]  Exponential  decay  ($d_i  =  \theta^i$,  $0  \leq  \theta  <  1$):
\begin{align}
V(\Lambda_T)  \sim  T  \quad  (T  \to  \infty).
\end{align}
The  process  exhibits  normal  diffusion  for  all  $\theta  <  1$;  
there  is  no  phase  transition.
\item[(ii)]  Power  decay  ($d_i  =  1/(i+1)^\gamma$,  $\gamma  \geq  0$):
\begin{align}
\label{scale}
V(\Lambda_T)  \sim  
\begin{cases}
T  &  (\gamma  >  1),  \\
T\log  T  &  (\gamma  =  1),  \\
T^{2-\gamma}  &  (0  \leq  \gamma  <  1).
\end{cases}
\end{align}
A  phase  transition  occurs  at  $\gamma  =  1$,  
separating  normal  diffusion  ($\gamma  >  1$)  
from  anomalous  super-normal  diffusion  ($\gamma  \leq  1$).
\end{enumerate}
\end{proposition}
%%%%%%%%%%%%%%%%%%
\begin{proof}
The  result  follows  from  Eq.~(\ref{V1})  together  with  
the  asymptotic  estimates  of  
$\sum_{i=1}^{T-1}(i+1)^{-\gamma}(T-i)$  
as  $T  \to  \infty$,  derived  in  
Eqs.~(\ref{g1})--(\ref{g2}).
\end{proof}
%%%%%%%%%%%%%%%%%%
\begin{remark}
This  proposition  establishes  the  super-normal  phase  
transition  as  a  genuine  feature  of  the  LGCP,  
manifesting  in  the  anomalous  scaling  of  the  
cumulative  variance.
The  transition  at  $\gamma  =  1$  is  a  property  of  
the  LGCP  itself  and  does  not  depend  on  the  
Merton  model  structure;  it  holds  for  any  LGCP  
with  power-law  temporal  correlations.
However,  in  the  present  paper  the  LGCP  is  derived  
from  a  Merton-type  credit  risk  model,  so  the  
transition  has  a  direct  interpretation:  
when  $\gamma  \leq  1$,  the  variance  of  cumulative  
defaults  grows  faster  than  linearly  in  $T$,  
implying  that  parameter  estimation  (e.g.,  of  
the  probability  of  default)  converges  more  slowly  
than  in  the  normal  case.
This  has  practical  consequences  for  credit  risk  
management,  as  more  data  is  required  to  achieve  
reliable  estimates.
While  the  super-normal  transition  is  known  in  the  
statistical  physics  literature  \cite{hod},  
its  appearance  in  LNCP  is  new.
\end{remark}

%%%%%%%%%%%%
\begin{proposition}[Persistence  of  impact]
\label{thm:impact}
Under  power  decay  with  exponent  $\gamma$,  
the  cumulative  log-impact  of  a  shock  $\delta$  
added  to  the  latent  variable  $\hat{y}_t$  satisfies
\begin{align}
\log  \frac{\bar{\lambda}_{\delta\infty}}{\bar{\lambda}_\infty}
=  \alpha\delta  \sum_{s=0}^{\infty}  \frac{1}{(s+1)^\gamma}.
\end{align}
The  impact  is  finite  when  $\gamma  >  1$  and  diverges  when  
$\gamma  \leq  1$,  consistent  with  the  phase  transition  
in  Proposition~\ref{thm:phase}.
Specifically,
\begin{enumerate}
\item[(i)]  $\gamma  >  1$:  
$\log(\bar{\lambda}_{\delta\infty}/\bar{\lambda}_\infty)  
<  \alpha\delta\bigl(1  +  1/(\gamma-1)\bigr)$  (finite).
\item[(ii)]  $\gamma  =  1$:  
$\log(\bar{\lambda}_{\delta\infty}/\bar{\lambda}_\infty)  
\sim  \alpha\delta\log  T$  (divergent).
\item[(iii)]  $\gamma  <  1$:  
$\log(\bar{\lambda}_{\delta\infty}/\bar{\lambda}_\infty)  
\sim  \alpha\delta  T^{1-\gamma}$  (divergent).
\end{enumerate}
\end{proposition}
%%%%%%%%%%%%%%%%%%
\begin{proof}
Using  the  cumulative  impact  measure  defined  in  Section  2.2.1.
The  result  follows  from  Eqs.~(\ref{i1})--(\ref{i2})  
together  with  the  asymptotic  estimate  of  
$\sum_{s=0}^{T-1}(s+1)^{-\gamma}$  as  $T\to\infty$,  
which  converges  for  $\gamma  >  1$  and  diverges  
as  $\log  T$  for  $\gamma  =  1$  and  as  $T^{1-\gamma}$  
for  $\gamma  <  1$.
\end{proof}

%%%%%%%%%%%%%
\begin{remark}
This  proposition  provides  a  complementary  view  of  
the  phase  transition  from  the  perspective  of  
impulse  response  analysis.
The  divergence  of  the  cumulative  log-impact  
when  $\gamma  \leq  1$  means  that  a  macroeconomic  
shock  at  time  $t$  continues  to  influence  default  
intensities  at  all  future  times,  with  the  total  
effect  growing  without  bound.
Note  that  the  mean  intensity  
$\bar{\lambda}  =  \lambda_0  e^{\alpha^2/2}$  
remains  finite  for  all  $\gamma$,  since  it  depends  
on  $\rho_A$  but  not  on  $\gamma$.
The  divergence  therefore  affects  the  
estimation  accuracy  of  parameters  
rather  than  the  stability  of  the  process  itself.
This  contrasts  with  the  Hawkes  and  SE-NBD  processes,  
in  which  the  mean  intensity  itself  diverges  
in  the  non-steady  phase  (see  Table~\ref{comp}).
\end{remark}
%%%%%%%%%%%%%%%%%%%
\noindent
These  three  propositions  together  establish  that  the  super-normal  
phase  transition  at  $\gamma  =  1$  is  a  genuine  feature  of  
the  LGCP,  
manifesting  both  in  the  scaling  of  the  cumulative  variance  
(Proposition~\ref{thm:phase})  and  in  the  persistence  of  
macroeconomic  shocks  (Proposition~\ref{thm:impact}).  
The  empirical  results  of  Section~\ref{4}  suggest  that  
pre-1980  credit  portfolios  operated  in  or  near  the  
super-normal  regime  ($\gamma  \leq  1$),  
whereas  more  recent  portfolios  are  typically  in  
the  normal  regime  ($\gamma  >  1$).

%%%%%%%%%%%%%%%%%%%%%%
\section{Concluding    Remarks}
This  study  considers  a  stochastic  point  process  with  
power-law  temporal  correlation  driven  by  hidden  variables.

We  show  that  a  generalized    Merton  type  model  under  exponential-tail  asset  
assumption---obtained  by  replacing  the  Gaussian  CDF  
with  a  logistic  CDF---converges  to  a  Log-Gaussian  Cox  
process  (LGCP)  with  log-normal  intensity  under  a  
double  scaling  limit.
The  essential  condition  for  this  convergence  is  not the Merton model structure per se, but hat  
the  asset  CDF  has  an  exponential  tail  
$\Phi(x)  \sim  Ae^{\kappa  x}$  as  $x  \to  -\infty$,
which  is  satisfied  by  the  logistic  and  generalized  
hyperbolic  distributions  but  not  by  the  Gaussian  
or  $t$-distributions  (see  Appendix~\ref{Ac}  and  Table~\ref{tab:tail}); 
the Merton model provides the specific factor
structure through which this tail condition translates into a double-scaling limit
and an LGCP with an economically interpretable intensity.
The LGCP can be a
useful tool for portfolio risk management in finance.

The  LGCP  can  be  a  useful  tool  for  portfolio  risk  
management  in  finance.

The  precise  role  of  the  Merton  model  structure  is  
as  follows.
The  factor  decomposition  in  Eq.~(\ref{G})  dictates  the  
double  scaling  limit  
$\lambda_0  =  Np'^{1/\sqrt{1-\rho_A}}$,
whose  exponent  $1/\sqrt{1-\rho_A}$  originates  from  
the  asset  correlation  $\rho_A$.
This  Merton-specific  scaling  is  what  allows  temporal  
correlations  to  survive  the  Poisson  limit:
under  the  standard  limit  $\lambda_0  =  Np$,  
the  dependence  on  $y_t$  vanishes  and  all  correlations  
are  lost.
Furthermore,  the  LGCP  parameter  
$\alpha  =  \kappa\sqrt{\rho_A/(1-\rho_A)}$
inherits  a  direct  interpretation  in  terms  of  
the  asset  correlation  $\rho_A$.
The  connection  between  the  Merton  model  and  the  LGCP  
via  this  double  scaling  limit  appears  to  be  new  
in  the  literature.

In  this  model  we  confirm  the  super-normal  transition  
when  the  temporal  correlation  follows  a  power  decay.
The  super-normal  transition  at  $\gamma  =  1$  and  the  
anomalous  scaling  $V(\Lambda_T)  \sim  T^{2-\gamma}$  
are  general  properties  of  the  LGCP  and  hold  for  any  
LGCP  with  power-law  temporal  correlations.
However,  since  the  LGCP  is  here  derived  from  a  
Merton-type  credit  risk  model,  the  transition  has  
a  direct  practical  interpretation:
when  $\gamma  \leq  1$,  the  variance  of  cumulative  
defaults  grows  faster  than  linearly  in  $T$,  
implying  that  parameter  estimation  of  the  
probability  of  default  converges  more  slowly  
than  in  the  normal  case.
This  is  an  important  issue  for  credit  risk  management:
when  estimating  PDs,  the  ranges  of  errors  are  wider  
than  in  the  normal  convergence  case,  and  more  data  
is  necessary  for  reliable  estimation.
For  the  case  of  exponential  decay,  there  is  no  
phase  transition.

We  discuss  the  relation  between  this  model  and  the  
Hawkes  and  SE-NBD  processes.
When  the  power  index  $\gamma$  is  larger  than  one,
the  PD  estimator  distribution  converges  normally.
One  can  see  that  the  Poisson  limit  inherits  the  
properties  of  the  discrete  model,  and  this  is  one  
of  the  reasons  for  the  importance  of  the  limit.
This  model  includes  two  correlations:  correlation  
in  the  same  time  period  and  temporal  correlation.

We  applied  this  model  to  default  portfolios.
We  found  that  the  power  decay  model  provides  better  
generalization  performance  for  long-term  data  with  
$\gamma  \leq  1$.
The  empirical  estimates  indicate  that  pre-1980  
portfolios  were  often  in  or  near  the  super-normal  
regime  ($\gamma  \leq  1$),  whereas  more  recent  
portfolios  are  typically  closer  to  the  normal  regime  
($\gamma  >  1$).
The  empirical  analysis  estimates  the  temporal  
correlation  parameter  $\gamma$  of  the  latent  
Gaussian  process  underlying  the  LGCP,  and  the  
finding  $\gamma  <  1$  for  pre-1980  portfolios  
provides  evidence  for  long-memory  behavior  
within  this  framework.
Thus,  the  theoretical  phase  transition  is  not  
merely  a  mathematical  artifact  but  appears  to  
have  practical  relevance  for  historical  credit  markets.

In  the  Hawkes  process  and  SE-NBD  process,  we  see  
a  phase  transition  that  is  not  the  super-normal  
transition.
In  the  limit,  the  power  distribution  appears  as  
the  intensity  function  at  the  critical  point.
Note  that  even  when  the  cumulative  log-impact  
diverges  ($\gamma  \leq  1$),  the  mean  intensity  
$\bar{\lambda}  =  \lambda_0  e^{\alpha^2/2}$  remains  
finite  for  all  $\gamma$,  since  it  depends  on  
$\rho_A$  but  not  on  $\gamma$.
This  contrasts  with  the  Hawkes  and  SE-NBD  processes,  
in  which  the  mean  intensity  itself  diverges  in  
the  non-steady  phase.
The  super-normal  transition  therefore  affects  
the  estimation  accuracy  of  parameters  rather  than  
the  stability  of  the  process  itself.

From  a  financial  perspective,  self-exciting  models  
can  be  viewed  as  representations  of  contagion  effects  
such  as  chain  bankruptcies,  whereas  the  LGCP  describes  
increases  in  defaults  driven  by  macroeconomic  conditions.
As  a  topic  for  future  research,  it  would  be  interesting  
to  investigate  what  types  of  phase  transitions  emerge  
when  these  two  mechanisms  are  combined,  and  which  
effect  is  more  dominant  in  empirical  data.
Furthermore,  a  direct  empirical  scaling  analysis  of  
$V(\Lambda_T)  \sim  T^{2-\gamma}$  would  provide  a  
more  direct  test  of  the  theoretical  phase  transition;
however,  this  requires  longer  time  series  than  
currently  available  and  is  left  for  future  work.

%%%%%%%%%%%%%%%%%%%%%%%%%%%%%%%%%%%%%%%%

\begin{appendices}

\section{Strong    correlation    limit    }
\label{Aa}

Here,    we    consider        the    obligors    that    are    independent    
of    the    economic    index    $y_t$.
The    PD    of    the    obligors    is    $p''$    and    the    distribution    
is    a    continuous    uniform    distribution    from    0    to    1    $\eta(0,1)$.
The    obligors    which    are    affected    by    the    economic    index    $y$    are    in    group    1    and    those    which    are    not    affected    are    in    group    2.
$G(y)$    corresponds    to    the    conditional    default    probability
as
\begin{equation}
G(y)=    \mbox{P}(X_{t}=1|y_{t}=y)=a\Phi    \left(\frac{Y-\sqrt{\rho_A}    y}{\sqrt{1-\rho_A}}\right)
+(1-a)p''z,
\end{equation}
where    the    ratio    of    group    1    is    $a$,    and    $z$    is    the    uniform    distribution.

The    distribution    of    the    number    of    the    default    is    
\begin{eqnarray}
P[X_t=k_t]&=&\int_{-\infty}^{\infty}    \frac{N!}{k_t!(N-k_t)!}
G(y)^{k_t}
(1-G(y))^{N-k_t}
\phi(y)    dy.
\end{eqnarray}
Here    we    use    the    cumulative    logistic    function    instead    of    a    normal        cumulative    distribution    as    Section  \ref{2},

in    the    case    $p\ll$1,    we    obtain,
\begin{equation}
    G(y)=    a    p'^{1/\sqrt{1-\rho_A}}    e^{-\frac{\sqrt{\rho_A}}{\sqrt{1-\rho_A}}\beta    y}+
    (1-a)    p''z,
\end{equation}
where    $y$    is    the    standard    normal    distribution,    and    $z$    is    the    uniform    distribution,    $z\sim    \rm{Uniform}(0,1)$

Here    we    take    the    limit
    $p'',    p',p\rightarrow    0$    and    $N    \rightarrow\infty    $    with    the    condition,    fixed
    $\lambda_1=N    p''$.
    Then,    we    obtain    
    \begin{equation}
\lambda(y,z)=a\lambda_0    e^{-\frac{\sqrt{\rho_A}}{\sqrt{1-\rho_A}}\beta    y}
+(1-a)\lambda_1    z
\end{equation}
The    expected    value    of    $\lambda(y)$    is
\begin{equation}
\bar{\lambda}=\int_{-\infty}^{\infty}\int_{0}^{1}
\lambda(y,z)\phi(y)\eta(z)    dy    dz
    =a\lambda_0    e^{\alpha^2/2}+(1-a)\lambda_1/2.
\end{equation}
This    process    can    be    written    as    follows,
\begin{equation}
    \lambda(\hat{y}_{t+1})=    a\lambda_0^{1-\theta}    e^{\sqrt{1-\theta^2}\alpha    \xi_{t+1}}\lambda(\hat{y}_t)^{\theta}+(1-a)\eta_t.    
\end{equation}
If    we    set    $\rho_A\rightarrow    1$    and    $\theta    \rightarrow    1$    with    fixed    $\sqrt{1-\theta^2}/\sqrt{1-\rho_A}=b$,    
It    is    the    Kesten    process    \cite{Kes},
\begin{equation}
    \lambda(\hat{y}_{t+1})=    a    e^{\beta    b    \xi_{t+1}}\lambda(\hat{y}_t)+(1-a)\eta_t.    
\end{equation}
With    the    condition    $\beta    b<1$,    it    is    the    power    distribution.
The    limit    is    the    strong    correlation    limit.
In    the    Hawkes        and    SE-NBD    processes,    we        observe    the    power    distribution    of    the    intensity    function    at    the    critical    point    of    the    phase    transition.
It    is    not    the    strong    correlation    limit.
Therefore,    the    temporal    correlation    of    the    self    excitation    is    stronger    than    that    of    hidden    variables.

\section{Details    of    Bayesian    Estimation}
\label{Ab}

The    estimation    procedure    adopted    in    this    study    is    outlined    below.

\subsection{Initial    Estimation    Without    Temporal    Correlation}

We    first    estimated    the    baseline    rate    $\lambda_0$    and    the    coefficient    $\alpha$    by    maximum    likelihood,    assuming    a    Poisson    model    without    temporal    correlation.    This    step    provides    initial    estimates        for    the        subsequent    procedures.

\subsection{Latent    Variable    Estimation    and    ACF    Evaluation}

Using    the    estimates    $\hat{\lambda}_0$    and    $\hat{\alpha}$,    we    inferred    the    latent    variable    sequence    $y_t$    by    maximizing    the    likelihood    under    the    non-correlated    model.    The    autocorrelation    function    (ACF)    of    the    inferred    $y_t$    series    was    then    computed    and    analyzed.

\subsection{Estimation    of    Decay    Parameters    via    ACF    Fitting}

The    ACF    of    $y_t$        can    be        fitted    to    two    functional    forms    using    least    squares:

\begin{itemize}
        \item    Exponential    decay:    $ACF(h)    \approx    \exp(-h/\theta)$
        \item    Power    decay:    $ACF(h)    \approx    (1    +    h)^{-\gamma}$
\end{itemize}

This    yields    the    estimates    of    $\theta$    and    $\gamma$    used    in    the    next    stage.

\subsection{Prior    Distribution    for    Bayesian    Estimation}

The    parameter    estimates    obtained    above    were    used    to    construct    normal    prior    distributions    in    the    following    Bayesian    model:

\[
\text{Parameter}    \sim    \mathcal{N}(\text{Estimate},    \,    (sc    \times    \text{Standard    Error})^2)
\]

Parameter    constraints    were    imposed    in    accordance    with    the    Stan    model    declarations:

\subsection{Selection    of    the    Prior    Scale    Factor    $sc$}

To    determine    the    appropriate    scale    factor    $sc$    for    the    prior    distributions,    the    predictive    performance    was    evaluated    for    each    candidate    value    using    Leave-Future-Out    cross-validation    (LFO-CV):

\[
sc    \in    \{1,    2,    3,    4,    5,    10,    20\}
\]

The    LFO-CV    procedure        systematically    varied        the    initial    prediction    point    $t_0$    in        each    dataset.    Specifically:

\begin{itemize}
        \item    For    the    first    three    datasets,    $t_0$    was    varied    from    50    to    100    in    steps    of    5:    $t_0    \in    \{50,    55,    60,    \dots,    100\}$.
        \item    For    the    remaining    six    datasets,    $t_0$    was    varied    from    30    to    42    in    unit    steps:    $t_0    \in    \{30,    31,    \dots,    42\}$.
\end{itemize}

For    each    combination    of    dataset,    model,    and    $sc$,    the    Bayesian    model    was    estimated,    and    the    predictive    performance    on    future    data    points    was    assessed    by    computing    the    LFO    score    across    the    range    of    $t_0$.    The    value    of    $sc$    yielding    the    lowest    total    LFO    score    was    selected    for    final    estimation.

\subsection{Bayesian    Estimation    for    Each    Dataset}

After    selecting    the    optimal    $sc$,    Bayesian    estimation    was    performed    on        each    dataset.    The    following    model    selection    and    diagnostic    criteria    were    computed:

\begin{itemize}
        \item    WAIC    (Widely    Applicable    Information    Criterion)
        \item    WBIC    (Widely    applicable    Bayesian    Information    Criterion)
        \item    LFO    (Leave-Future-Out    Information    Criterion)
\end{itemize}

Note    that    LFO    (Leave-Future-Out)    cross-validation    is    more    suitable    than    WAIC    for    time-series    data,    because        it    evaluates    the    model's    ability    to    predict    future    values    based    on    past    observations    while    accounting    for    temporal    dependencies.

The    convergence    of    the    Markov    chains    was    assessed    using    the    Gelman-Rubin    statistic    $\hat{R}$.    All    outputs,    including    parameter    estimates,    information    criteria,    and    $\hat{R}$    values,    were    recorded    in    data    frames    for    further    analysis    and    reporting.

The    R-code    for        this    study    is    available    at    GitHub\cite{Git}.

Disclosure    of    interest

The    authors    declare    no    competing    interests.

Data    Availability    Statement

The    data    that    support    the    findings    of    this    study    are    publicly    available    from    \cite{Data1,Data2}.    

\section{Tail    distribution    and    the    limit}
\label{Ac}
In    section    \ref{2}    we    use    the    logistic    function    instead    of    the    cumulative    normal    distribution.
In    this    Appendix    we    clarify    the    origin    of    the    log    normal    intensity    function.
In    the    Poisson        limit    we    use        not        the    whole    logistic    function    but    its    tail.        
We    consider    the    exponential    decay        at    the    tail    for    the    cumulative    normal    distribution,
\begin{equation}
\Phi(x)\sim    A    e^{\kappa    x}
\end{equation}
when    $x$    is    sufficiently    small,    $x\sim    -\infty$.
$A$    and    $\kappa$    are    the    constants.
Using    this    distribution,    we    obtain
\begin{equation}
G(y)=A^{1-1/\sqrt{1-\rho_A}}    p'^{1/\sqrt{1-\rho_A}}    e^{-\frac{\sqrt{\rho_A}}{\sqrt{1-\rho_A}}\kappa    y}.
\end{equation}
It    is    consistent    with    Eq.(\ref{G}).
Therefore,    we    can    confirm        the    origin    of    the    log    normal    intensity    function    is    the    exponential    decay    at    the    tail.    
%Another    distribution    which    have    the    exponential    decay    at    the    tail    is    the    Laplace    distribution.
In    fact,        some    extensions    of    the    Merton    model        have    fat    tail    distributions,    because    the    distribution    of    assets    have    the    fat    tail  \cite{HW,Zho,Dem,Fre,Sch2}.    In        \cite{HW}    $t$    distribution        is    used        for    the    Merton    model.    
The    tail    of        $t$    distribution        is    a    power.
In  \cite{Fre}  generalized  hyperbolic  distribution  is  used  for  the  Merton  model.  The  tail  is  same  as  logistic  function.

It    is    important    to    note,        that    this    derivation    relies    on    an    exponential    tail    behavior.
The    cumulative    normal    distribution        does    not    exhibit    a    simple    exponential    tail:    for    large    negative    $x$    we    have        asymptotic
\[
\Phi    (x)\sim        \frac{1    }    {\sqrt{2\pi}    x}    \exp    (-x^2/2),
\]
as    $x\rightarrow    -\infty$,    so    that    the    decay    is    $\exp    (-x^2/2)    $    rather    than    purely    exponential.
The    logistic    function    has    heavier        tail    than    the    normal        distribution.
The    assumption,    the    use    of    the    logistic    function    instead    of    the        normal    distribution    is    intended    to    capture    the    fat    tails    of    the        asset    distribution    \cite{Man}.
When    the        decay    is    $\exp    (-x^2/2)    $,    it    is    difficult    to    take    the    Poisson    limit.
We  summarized  in  Table  \ref{tab:tail}  the  limit  of  several  Merton  models.
\begin{table}[h]
\centering
\caption{
Tail  behavior  of  representative  asset  CDFs  
$\Phi(x)$  as  $x  \to  -\infty$    of  generalized  Merton  models  and  whether  the  
LGCP  limit  is  obtained  under  the  double  scaling  limit.
Here  $\beta$,  $\alpha$,  $\delta$,  $\nu$  are  
distribution  parameters.
}
\label{tab:tail}
\begin{tabular}{llcc}
\hline
Asset  Distribution  &  Tail  behavior  of  $\Phi(x)$  
&  Exponential  tail?  &  LGCP  limit?  \\
\hline
Gaussian  (normal)  \cite{Mer}
&  $\sim  \frac{1}{\sqrt{2\pi}|x|}e^{-x^2/2}$  
&  No  &  No  \\
Logistic  
&  $\sim  e^{\beta  x}$  
&  Yes  ($\kappa  =  \beta$)  &  Yes  \\
Generalized  hyperbolic    \cite{Fre}
&  $\sim  e^{-(\alpha-|\beta|)|x|}$  
&  Yes  ($\kappa  =  \alpha-|\beta|$)  &  Yes  \\
$t$-distribution  ($\nu$  d.f.)  \cite{HW}
&  $\sim  |x|^{-\nu}$  (power  law)  
&  No  &  No  \\
\hline
\end{tabular}
\end{table}

%%=============================================%%
%%  For  submissions  to  Nature  Portfolio  Journals  %%
%%  please  use  the  heading  ``Extended  Data''.      %%
%%=============================================%%

%%=============================================================%%
%%  Sample  for  another  appendix  section			              %%
%%=============================================================%%

%%  \section{Example  of  another  appendix  section}\label{secA2}%
%%  Appendices  may  be  used  for  helpful,  supporting  or  essential  material  that  would  otherwise  
%%  clutter,  break  up  or  be  distracting  to  the  text.  Appendices  can  consist  of  sections,  figures,  
%%  tables  and  equations  etc.
\end{appendices}

%%===========================================================================================%%
%%  If  you  are  submitting  to  one  of  the  Nature  Portfolio  journals,  using  the  eJP  submission      %%
%%  system,  please  include  the  references  within  the  manuscript  file  itself.  You  may  do  this    %%
%%  by  copying  the  reference  list  from  your  .bbl  file,  paste  it  into  the  main  manuscript  .tex  %%
%%  file,  and  delete  the  associated  \verb+\bibliography+  commands.                                                        %%
%%===========================================================================================%%

%\bibliography{sn-bibliography}%  common  bib  file
%%  if  required,  the  content  of  .bbl  file  can  be  included  here  once  bbl  is  generated
%%\input  sn-article.bbl
\noindent
\textbf{  Funding}
This  work  was  supported  by  the  Japan  Society  for  the  Promotion  of  Science  (JSPS)  KAKENHI  Grant  Number  26K06955  (Grant-in-Aid  for  Scientific  Research  (C)).

\end{document}